\documentclass[aps,prb,twocolumn,10pt,longbibliography,floatfix,superscriptaddress]{revtex4-2}

\usepackage{physics}
\usepackage[english]{babel}
\usepackage[utf8]{inputenc}
\usepackage{amssymb}
\usepackage{amsmath}
\usepackage{dsfont}
\usepackage{multirow}

\usepackage[pdftex]{graphicx}
\usepackage{xspace}
\usepackage{dsfont}
\usepackage{bm}
\usepackage[pdfencoding=auto, psdextra]{hyperref}
\hypersetup{
    colorlinks,%
    citecolor=blue,%
    filecolor=blue,%
    linkcolor=blue,%
    urlcolor=blue
}

\usepackage[usenames, dvipsnames]{xcolor}

\begin{document}

\title{Interplay between boundary conditions and Wilson's mass in Dirac-like Hamiltonians}
\author{A. L. Araújo}
\author{R. P. Maciel}
\author{R. G. F. Dornelas} 
\affiliation{Instituto de Física, Universidade Federal de Uberlândia, Uberlândia, MG 38400-902, Brazil}

\author{D. Varjas}
\affiliation{QuTech and Kavli Institute of Nanoscience, Delft University of Technology, P.O. Box 4056, 2600 GA Delft, The Netherlands}

\author{G. J. Ferreira}
\affiliation{Instituto de Física, Universidade Federal de Uberlândia, Uberlândia, MG 38400-902, Brazil}
\date{\today}

\begin{abstract}
Dirac-like Hamiltonians, linear in momentum $k$, describe the low-energy physics of a large set of novel materials, including graphene, topological insulators, and Weyl fermions. We show here that the inclusion of a minimal $k^2$ Wilson's mass correction improves the models and allows for systematic derivations of appropriate boundary conditions for the envelope functions on finite systems. Considering only Wilson's masses allowed by symmetry, we show that the $k^2$ corrections are equivalent to Berry-Mondragon's discontinuous boundary conditions. This allows for simple numerical implementations of regularized Dirac models on a lattice, while properly accounting for the desired boundary condition. We apply our results on graphene nanoribbons (zigzag and armchair), and on a PbSe monolayer (topological crystalline insulator). For graphene, we find generalized Brey-Fertig boundary conditions, which correctly describe the small gap seen on \textit{ab initio} data for the metallic armchair nanoribbon. On PbSe, we show how our approach can be used to find spin-orbital-coupled boundary conditions. Overall, our discussions are set on a generic model that can be easily generalized for any Dirac-like Hamiltonian.
\end{abstract}

\maketitle

\section{Introduction}

The Dirac-like Hamiltonians play an ubiquitous role in novel materials, ranging from graphene \cite{Graphene1, Graphene2, CastroNeto2009Review} to topological insulators (TI) \cite{bernevig2006BHZ, Bernevig2006QSHE, Kane2010Review, Zhang2011Review, shen2012topological, bernevig2013book}, its crystalline \cite{Fu2011TCI, SnTe1, Slager2013SGClass, PbSe1,  Fu2015ReviewTCI, Slager2017TopoClassCombina} and higher-order \cite{Benalcazar2017Multipole, Benalcazar2017QEML, Langbehn2017HOTI, Schindler2018HOTI} TI counterparts, and Weyl semimetals \cite{Wan2011WSM, Xu2015WSMExp}. The Dirac cone structure of their low-energy band dispersion leads to great interest for possible optoelectronic and spintronic devices \cite{device1,device2,device4,device3}. Since the Dirac cone itself is well described by linear in momentum $k$ Hamiltonians, bulk models could be limited to this leading-order contribution. However, the $k^2$ Wilson's terms \cite{wilson1974confinement, Kogut1975, NIELSEN1981219, nielsen1981absence} are required to regularize the models for the calculation of topological invariants \cite{shen2012topological, bernevig2013book, Denis2018Paradoxical}. Moreover, numerical (finite-differences) implementations of $k$-linear models face the fermion-doubling problem \cite{Beenakker2008FiniteDiff, AlexisCaio2012, zhou2016LatticeModel}. For finite systems (\textit{e.g.}, nanoribbons), the Dirac models allow for a variety of possible non-trivial boundary conditions \cite{Peeters2011CircularGrapheneDot, Peeters2011GrapheneDots, Peeters2016ABCgraphene}, depending on the broken symmetry that imposes the confinement \cite{McCannFalko2004, Akhmerov2008Boundary}, as initially discussed by Berry and Mondragon \cite{BerryMondragon1987}. However, if the $k^2$ corrections are included in the model, one expects that the only allowed boundary condition is that of a vanishing envelope function at the edges. Therefore, we ask the following: How can the different $k$-linear boundary conditions be translated to models that account for the $k^2$ corrections?

In this paper, we investigate this question to show that the proper choice of the $k^2$ Wilson's correction induces the desired boundary conditions on Dirac-like ($k$-linear) Hamiltonians. In the first part of the paper we establish our results on a generic model that applies for all Dirac-like materials. We formulate this discussion using group-theory arguments, thus emphasizing its generality, while providing a recipe on how to apply our ideas to different materials. 
For the Bernevig-Hughes-Zhang (BHZ) model \cite{bernevig2006BHZ}, the relation between the hard-wall boundary condition and the $k^2$ terms was shown in Ref.~\onlinecite{Denis2018Paradoxical}. Here, we apply our method to the well-known graphene zigzag and armchair nanoribbons, and to the PbSe monolayer topological crystalline insulator. Indeed, graphene is an ideal material to test our findings due to (i) the formation of edge-state bands connecting the $K$ and $K'$ valleys on the zigzag case; and (ii) the contrast between gapped and gapless dispersions of armchair nanoribbons with different widths. We show that our systematic approach allows us to derive and generalize the Brey and Fertig boundary conditions \cite{BreyFertig2006} from symmetry constraints, which is more general than the usual analysis of the atomistic terminations, thus extending the derivation of boundary conditions to naturally include spinful systems (\textit{e.g.}, PbSe and SnTe TCIs).

In a previous paper \cite{Bruno2017FDP}, our group has shown that the $k^2$ Wilson's mass term \cite{wilson1974confinement} allows for a simple elimination of the numerical fermion-doubling problem on finite-differences implementations \cite{Beenakker2008FiniteDiff, AlexisCaio2012, zhou2016LatticeModel}. A similar proposal is established in Ref.~\cite{zhou2016LatticeModel}, however, their choice of Wilson's term undesirably breaks time-reversal symmetry. In Ref.~\cite{Bruno2017FDP}, it was suggested, and shown as a conjecture, that the proper choice of the $k^2$ term avoids this undesired broken symmetry. Here, we prove this conjecture and extend it to show how it can be used to either (i) derive non trivial boundary conditions for Dirac-like materials on $k$-linear models, or (ii) properly regularize the Dirac-models on a lattice by choosing the appropriate $k^2$ Wilson's term that accounts for the desired type of boundary condition.

The band structures from the effective Hamiltonians are compared with \textit{ab initio} results obtained from density functional theory (DFT). The effective Hamiltonians are obtained with support from QSYMM python's package \cite{Qsymm2018}, and the tight-binding models are implemented with KWANT python's package \cite{kwant}. All codes, input, and data files are available as Supplemental Material \cite{SM}. For the DFT simulations, we use the generalized gradient approximation (GGA) for the exchange and correlation functional \cite{GGAPBE35}. Fully relativistic $j$-dependent pseudopotential, within the projector augmented wave method \cite{PAW36}, was used in the noncollinear spin-DFT formalism self-consistently. We use the Vienna \textit{ab initio} simulation package (VASP) \cite{VASP33,VASP34}, with plane wave basis set with a cut-off energy of 400--500 eV. The Brillouin zone is sampled using a number of $k$-points such that the total energy converges within the meV scale. The optimized force criteria for convergence was less than 0.01 eV/\AA.

\section{Generic model}

Effective models can be obtained from symmetry constraints imposed by method of invariants \cite{winkler2003spin}, which is equivalent to a $k\cdot p$ envelope function approach, yielding a matrix expansion of the Hamiltonian $H\equiv H(\bm{k})$ in powers of the momentum $\bm{k}$. For Dirac-like materials, one might truncate the expansion on the leading order ($k$-linear terms), for which the confinement is set by non-trivial boundary conditions \cite{BerryMondragon1987, McCannFalko2004, BreyFertig2006, Akhmerov2008Boundary, Bruno2017FDP} (see Sec. \ref{sec:hardwall}). However, the $k^2$ corrections play a significant role in numerical simulations, allowing for a simple elimination of the fermion-doubling problem \cite{Beenakker2008FiniteDiff, AlexisCaio2012, Bruno2017FDP}. Moreover, a conjecture introduced in Ref.~\onlinecite{Bruno2017FDP} states that the matrix form of the $k^2$ term is related to the hard-wall boundary conditions \cite{BerryMondragon1987}. In this section, we prove this conjecture on a generic, yet complete, formulation in terms of a minimal model.

To guide our discussions, let us consider a one-dimensional system given by the generic Hamiltonian
\begin{equation}
    H = \hbar v_F U_k k + m U_w k^2 + U_c V(x),
    \label{eq:genH}
\end{equation}
which is defined along $x$, and $k = -i\partial_x$. The $k$-linear term gives the Dirac-like dispersion at low energies with Fermi velocity $v_F$. The $k^2$ correction introduces the Wilson's mass $m$ \cite{wilson1974confinement, Bruno2017FDP}. The last term is a soft-wall confining potential given by a symmetric profile $V(x) = V_0[1-\Theta(x+L)+\Theta(x-L)]$, where $\Theta(x)$ is the Heaviside step function, and $V_0$ is the intensity. This profile defines the physical system within the inner region $|x|<L$, while on the outer region ($|x|\geq L$) it opens a gap $2|V_0|$. Later, we will consider the hard-wall limit $|V_0| \rightarrow \infty$, which excludes the outer region from the physical domain. The $U_k, U_w, U_c$ are Hermitian matrices defined by the symmetry constraints imposed on $H$, which will be discussed throughout the next sections. Namely, $U_k$ defines the kinetic energy term, $U_w$ sets the type Wilson's mass, and $U_c$ sets the form of confinement.

Hereafter, we assume that the $U_k$, $U_c$, and $U_w$ matrices are nonsingular. Neglecting the $k^2$ correction by setting $m=0$, $U_k$ defines the unbounded Dirac-like spectrum of the Hamiltonian, \textit{i.e.}, its eigenvalues give the positive and negative velocities of the Dirac cone. Thus, we require $\det(U_k) \neq 0$, otherwise, one would have a flat branch in the energy dispersion. In turn, the $U_c$ matrix defines the confinement gap on the outer region where $V(x) \neq 0$. Within this region, we demand that Eq.~\eqref{eq:genH} does not have any propagating modes at zero energy. To guarantee this, we demand that $U_k^{-1} U_c$ has no purely real eigenvalues, which implies $\det(U_c) \neq 0$. The anticommutation condition $\left\{U_k, U_c\right\} = 0$ is sufficient, but not necessary, to fulfill this. A similar argument applies to the case with a Wilson's mass term in the interior of the sample, where $V(x) = 0$ and $m \neq 0$. We demand that the Wilson's mass term does not introduce any new propagating modes at zero energy beyond the ones in the original Dirac model. This results in the identical conditions for $U_w$ as for $U_c$ (for details, see Appendix~\ref{app:rigorous}).

\subsection{Symmetry constraints: \texorpdfstring{$U_w \equiv U_c$}{Uw = Uc}}
\label{sec:UwequalUc}

Let us consider that our generic system, i.e., $H$ from Eq.~\eqref{eq:genH}, is invariant under a symmetry group $\mathcal{G}$, which is composed by two types of symmetry operations: $\mathcal{S}_+$ and $\mathcal{S}_-$. The $\mathcal{S}_+$ operators leave $x$ invariant, while $\mathcal{S}_-$ takes $x\rightarrow -x$. Therefore, it follows the transformations $\mathcal{S}_\pm x \mathcal{S}_\pm^{-1} = \pm x$, $\mathcal{S}_\pm k \mathcal{S}_\pm^{-1} = \pm k$, $\mathcal{S}_\pm V(x) \mathcal{S}_\pm^{-1} = V(x)$. The last one is a consequence of the symmetric form of $V(x)$ introduced previously. Imposing that $H$ is invariant over the full group $\mathcal{G}$, (\textit{i.e.}, $\mathcal{S}_\pm H \mathcal{S}_\pm^{-1} = H$), one obtains the symmetry constraints for the matrices of $H$:
\begin{align}
    [U_w, D^\psi(\mathcal{S}_\pm)] &= [U_c, D^\psi(\mathcal{S}_\pm)] = 0,
    \label{eq:UwUc}
    \\
    [U_k, D^\psi(\mathcal{S}_+)] &= \{U_k, D^\psi(\mathcal{S}_-)\} = 0,
\end{align}
which defines the symmetry-allowed matrices $U_k$, $U_w$, and $U_c$. Here $[\cdot,\cdot]$ and $\{\cdot,\cdot\}$ are the commutator and anti-commutator operations, and $D^\psi(\mathcal{S}_\pm)$ are the matrix representations of $\mathcal{S}_\pm$ in the Hilbert space. Since $U_w$ and $U_c$ satisfy the same constraints, it follows that they are equivalent ($U_w \equiv U_c$), \textit{i.e.}, both are in the same linear space of allowed matrices. This equivalence between $U_w$ and $U_c$ was assumed truthful, but not rigorously proven in Ref.~\onlinecite{Bruno2017FDP}.

\subsection{Hard-wall boundary conditions}
\label{sec:hardwall}

The appropriate hard-wall boundary condition depends on the order of the differential equation. For our generic effective model $H$ in Eq.~\eqref{eq:genH}, the Schrödinger equation $H F(x) = E F(x)$ has order 2 if $m\neq 0$, or order 1 if $m=0$. Here, $F(x)$ is an envelope spinor function \cite{winkler2003spin}. In all cases, the energy $E = \bra{F}H\ket{F}$ must be bounded and well defined. Consider $H$ from Eq.~\eqref{eq:genH}, with a simplified single boundary profile $V(x) = V_0 \Theta(x)$. On the outer region $x>0$ the gap $|2V_0|$ yields evanescent solutions at low energies, \textit{i.e.}, $F(x>0) \sim F_0 e^{-x/\lambda}$, where $F_0$ is the spinor at $x=0$. If the $k$-linear term dominates the low-energy band structure, the penetration length is $\lambda \propto \hbar v_F/|V_0|$. In the hard-wall limit $|V_0|\rightarrow \infty$ and $\lambda \rightarrow 0$, thus near the interface $x\approx 0$ we can write $F(x) \approx F_0 [1-\Theta(x)]$. Considering only the $x\approx 0$ range on the integrals in $E = \bra{F}H\ket{F}$, it can be shown that the contributions from the $k$-linear and potential $V(x)$ terms are finite, while $m\bra{F}U_w k^2\ket{F} \approx -m F_0^\dagger U_w F_0 \delta(0)$ is ill defined due to the $\delta(0)$. Therefore, either $m=0$ or $F_0=0$. In the first case, one gets a $k$-linear model with discontinuous $F(x)$, while the second case yields a $k^2$ model with a continuous $F(x)$ that vanishes at the hard-wall interface.

The discontinuous behavior of $F(x)$ in $k$-linear Hamiltonians ($m=0$) was first introduced in the neutrino billiards by Berry and Mondragon (BM)~\cite{BerryMondragon1987}, and further discussed in Refs.~\onlinecite{Alonso1997DiracBC2, Alonso1997DiracBC}. Later, it was applied to graphene \cite{McCannFalko2004, Akhmerov2008Boundary} and topological insulators~\cite{Ferreira2013Magnetically, Bruno2017FDP, Denis2018Paradoxical}. Here, we cast the BM hard-wall boundary condition for $H$ in Eq.~\eqref{eq:genH} in a form that explicitly shows $U_k$ and $U_c$ as
\begin{align}
    \left( i U_k^{-1} U_c \right) F_0 &= \frac{\hbar v_F}{\lambda V_0} F_0.
    \label{eq:linearBC}
\end{align}
This is an eigenvalue equation for the matrix $i U_k^{-1} U_c$ with eigenvalue $\alpha = \hbar v_F / \left(\lambda V_0 \right)$. The solutions with $\Re \lambda > 0$ ($\Re \alpha > 0$) describe states decaying outside the boundary. Defining $M = \alpha^{-1} \left( i U_k^{-1} U_c \right)$, the boundary condition becomes
\begin{align}
    M F_0 &= F_0.
    \label{eq:BM}
\end{align}
If $U_c$ and $U_k$ are chosen such that $M$ is traceless, unitary, and Hermitian, this reproduces the most general current-conserving boundary conditions~\cite{BerryMondragon1987, McCannFalko2004, Akhmerov2008Boundary}.
We derive formally the same boundary condition replacing $U_c$ with $U_w$ for the case of a hard wall with Wilson's mass term (for details, see Appendix~\ref{app:rigorous}).
In the remaining of the paper we choose the Hamiltonian terms such that $\left( i U_k^{-1} U_c \right)$ has at least one pair of eigenvalues $\alpha = \pm 1$, equivalently $\det \left(U_k \mp iU_c \right) = 0$.
Using this, we cast the boundary conditions Eq.~\eqref{eq:BM} for a finite system with $|x| < L$ as $(U_k \mp iU_c)F(\pm L) = 0$.

\subsection{Summary of the models}

From the considerations above, we conclude that one can choose to work with either the $k$-linear formulation with nontrivial boundary conditions, or the $k^2$ model with trivial hard-walls:
\begin{align}
    H = \hbar v_F U_k k, 
        \quad\quad\quad\text{ with }
    (U_k \mp i U_c)F(\pm L) &= 0,
    \label{eq:Hk1}
    \\
    H = \hbar v_F U_k k + m U_w k^2, 
        \quad\quad\quad\text{ with }
    F(\pm L) &= 0.
    \label{eq:Hk2}
\end{align}
These two approaches are equivalent, due to the analytical properties of the boundary conditions discussed above, and the equivalence $U_w \equiv U_c$ shown in Sec.~\ref{sec:UwequalUc}. Thus, for the $k$-linear model of Eq.~\eqref{eq:Hk1}, the characteristics of the confinement appear on the BM boundary condition, while on the $k^2$ model of Eq.~\eqref{eq:Hk2}, it enters through Wilson's $k^2$ term.
For the $k^2$ model, the Wilson's mass can be set within
\begin{equation}
    \dfrac{\delta_x^2\delta_\varepsilon}{2} \leq |m| \leq \dfrac{(\hbar v_F)^2}{\delta_\varepsilon},
    \label{eq:mass}
\end{equation}
which ensures that the $k$-linear term dominates the low-energy spectrum within the $\delta_\varepsilon$ energy range and eliminates the fermion doubling problem on a lattice \cite{Bruno2017FDP}. In the analytical limit, the pure Dirac-like model is restored as $m\rightarrow 0$~\cite{massRange}.

We explicitly show in Appendix~\ref{app:rigorous} that the envelope functions of the two models match far from the boundaries. Near the boundary, the $k^2$ model has an additional localized contribution that decays quickly away from the boundary in the limit of small $m$~\cite{massRange}.

\section{Graphene nanoribbons}

\begin{figure}[t]
    \centering
    \includegraphics[width=\columnwidth]{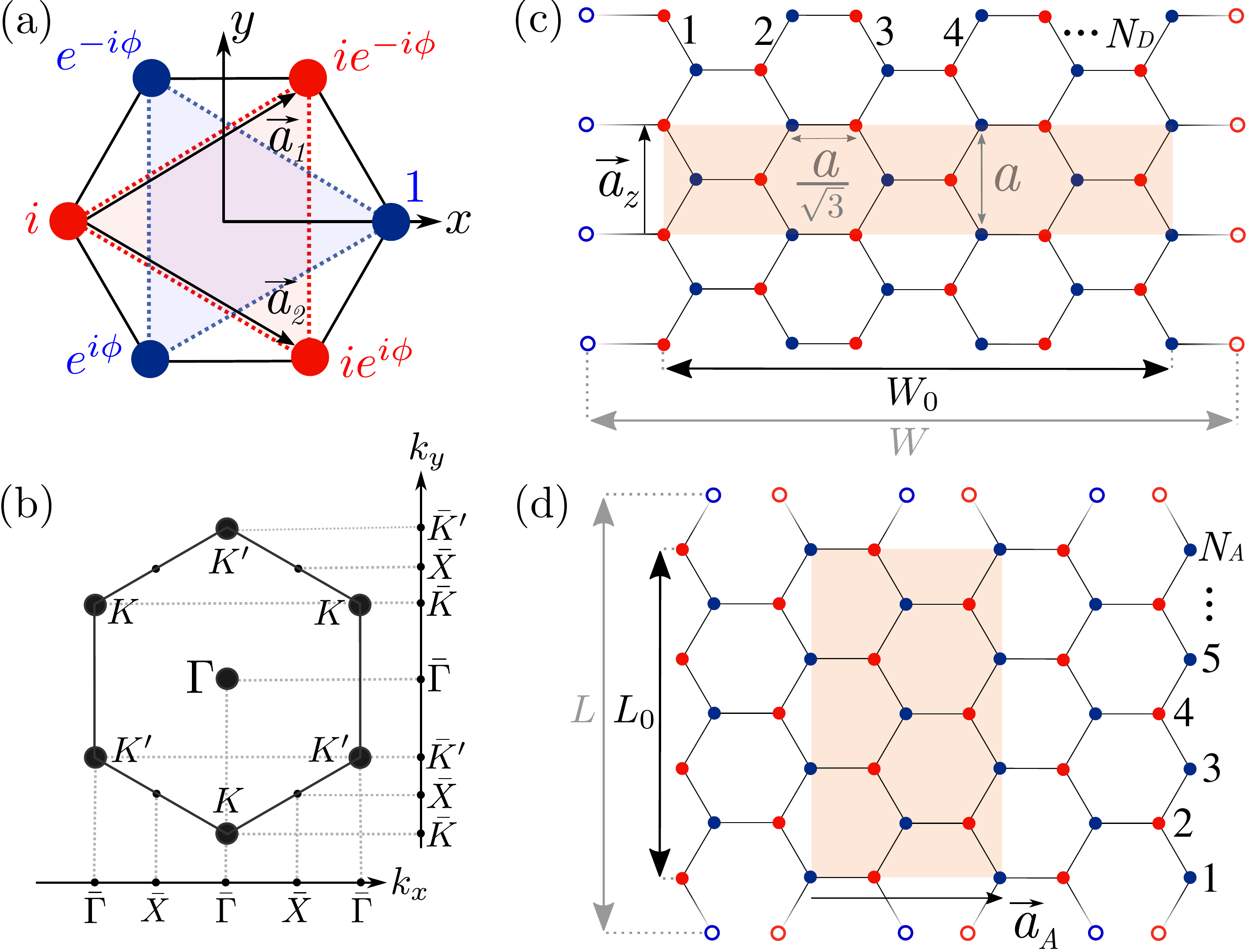}
    \caption{(a) Hexagonal cell of monolayer graphene with the $A$ (blue) and $B$ (red) sublattices emphasized. The $K$-point basis functions $\varphi_\mu(\bm{r})$ (with $\mu=A, B$) for each sublattice is composed by $p_z$ orbitals multiplied by the Bloch phases indicated at each site, with $\phi=2\pi/3$. For $K'$, $\varphi_{\mu'}(\bm{r})$ are composed replacing $\phi \rightarrow -\phi$. (b) Projections of the $K$, $K'$, and $\Gamma$ points of the Brillouin zone into its one-dimensional counterparts for zigzag ($k_y$) and armchair ($k_x$) nanoribbons. Lattices for (c) zigzag and (d) armchair nanoribbons with $N_D$ ($N_A$) dimers (atoms) from edge to edge. Their primitive vectors $\bm{a}_{Z/A}$ and unit cells are highlighted. The carbon-carbon distance is $a/\sqrt{3} \approx 0.142$~nm. The empty circles at the edges indicate the lattice sites that were removed to form each nanoribbon, defining the effective dimensions $W = W_0 + 2a/\sqrt{3}$ and $L = L_0 + a$, where $W_0 = (3N_D/2 - 1)a/\sqrt{3}$ and $L_0 = (N_A - 1)a/2$.}
    \label{fig:GrapheneRibbons}
\end{figure}

Graphene nanoribbons (Fig.~\ref{fig:GrapheneRibbons}) are ideal cases to present our findings on a concrete system, since its band structure and boundary conditions are well known \cite{BreyFertig2006, CastroNeto2009Review}. In Ref.~\onlinecite{BreyFertig2006}, Brey and Fertig (BF) have shown that the graphene $k$-linear boundary conditions depend on the nanoribbon atomic termination. Particularly, their boundary condition predicts a metallic armchair nanoribbon with an identically zero gap for $N_A = 3p+2$ (with integer $p$), while the \textit{ab initio} data from Ref.~\onlinecite{Son2006DFTgap} shows a vanishing, but finite gap. Indeed, the null gap is not consistent with its symmetry group ($\sim D_{2h}$, see Appendix \ref{app:fullk2graphene}) of the nanoribbons. 

In this section we systematically revise the BF boundary conditions. We find that it is equivalent to the BM boundary condition, given by a proper choice of $U_c$, which is imposed by symmetry \cite{McCannFalko2004, Akhmerov2008Boundary}. However, for the metallic armchair case, we find that $U_c$ diverges, which is a consequence of the zero gap inconsistency mentioned above. Therefore, we propose a generalization of the BF boundary condition that fixes this inconsistency in both $k$-linear and $k^2$ approaches.

Initially, let us consider the usual $k$-linear Dirac model of a full monolayer graphene. Later in Section \ref{sec:k2graphene}, we introduce the $k^2$ model for the nanoribbons. The $k\cdot p$ expansion for graphene considers basis functions given by its solutions at $K$ and $K'$ valleys, \textit{i.e.,} $\varphi_{A}(\bm{r})$, $\varphi_{B}(\bm{r})$, $\varphi_{A'}(\bm{r})$ and $\varphi_{B'}(\bm{r})$, where $A$ and $B$ label the sublattices. These are illustrated in Fig.~\ref{fig:GrapheneRibbons}(a). Within the envelope function approximation \cite{DiVincenzo1984GrapheneKP, BastardBook, Ando2005ReviewNanotubes}, the expansion reads as
\begin{align}
    \psi(\bm{r}) = \sum_{\mu} \Big[
    f_\mu(\bm{r})e^{i\bm{q}\cdot\bm{r}}\varphi_\mu(\bm{r})
    +
    f_{\mu'}(\bm{r})e^{i\bm{q}'\cdot\bm{r}}\varphi_{\mu'}(\bm{r})\Big],
    \label{eq:psiexp}
\end{align}
where $f_\mu(\bm{r})$ and $f_{\mu'}(\bm{r})$ are the envelope functions, $\mu=\{A,B\}$ label the sublattices, $\bm{q} = \bm{k} - \bm{K}$ and $\bm{q}' = \bm{k} - \bm{K}'$ are the deviations from the $K$ and $K'$ valleys in $k$-space. Bloch theorem requires $\psi(\bm{r}+\bm{R})  = e^{i\bm{k}\cdot\bm{R}}\psi(\bm{r})$, where $\bm{R} = n_1 \bm{a}_1 + n_2 \bm{a}_2$ (with $n_1$, $n_2$ integers) is a Bravais translation of the monolayer. Since the Bloch phase in $\varphi_{\mu^{(\prime)}}(\bm{r}+\bm{R}) = e^{i\bm{K}^{(\prime)}\cdot\bm{R}}\varphi_{\mu^{(\prime)}}(\bm{r})$ cancels out the opposite phase in the $q$-exponentials, Bloch theorem is satisfied for a periodic $f_{\mu^{(\prime)}}(\bm{r}+\bm{R}) = f_{\mu^{(\prime)}}(\bm{r})$. Up to leading order in $\bm{k}$, the usual effective Dirac-like Hamiltonian is 
\begin{align}
    H_G(\bm{k}) &= h(\bm{q}) \oplus h^*(\bm{q}'),
    \label{eq:hq}    
    \\
    h(\bm{q}) &= \hbar v_F \bm{\sigma}\cdot\bm{q}.
\end{align}
The $4\times 4$ $H_G(\bm{k})$ Hamiltonian acts on the envelope spinor $F(\bm{r}) = [ f_A(\bm{r}), f_B(\bm{r}), f_{A'}(\bm{r}), f_{B'}(\bm{r}) ]$. Notice that we write $H_G(\bm{k})$ in terms of the deviations $\bm{q}$ and $\bm{q}'$, such that the Dirac cones occur at $\bm{k} \sim \bm{K}$ and $\bm{K}'$. This notation will be useful to keep track of the nanoribbon confinement projections onto the $k_x$ (armchair) or $k_y$ (zigzag) axis in the next sections. There, the projections will retain the overall form of the $\psi(\bm{r})$ expansion above, but they will change the definitions of $\bm{q}$ and $\bm{q}'$.

\subsection{Revised Brey and Fertig boundary conditions}
\label{sec:revisedBF}

An elegant approach to the boundary conditions for graphene nanoribbons was introduced in Ref.~\cite{BreyFertig2006} by Brey and Fertig. There, they propose that the envelope function must vanish at the sites that were removed to form the nanoribbons. Consequently, it depends on the atomic terminations, rendering different boundary conditions for the zigzag and armchair cases. We have recently used this approach to obtain boundary conditions for topological crystalline insulators \cite{Araujo2016Nonsymm}.

Next, we revise and generalize these boundary conditions for zigzag and armchair nanoribbons, and in the next section we show their equivalence to the $U_k$ and $U_c$ matrices on the BM approach. Complementary, the boundary conditions for confinement in arbitrary directions (beyond the zigzag and armchair) and different atomistic terminations were studied in Ref.~\cite{Akhmerov2008Boundary}. Their results can be used to define general $U_k$ and $U_c$ matrices.

\subsubsection{Zigzag nanoribbons}

To model the zigzag nanoribbons, we start from the bulk basis functions $\varphi_{\mu,\mu'}(\bm{r})$, but with a modified $\psi(\bm{r})$ expansion. Namely, replace $\bm{q}^{(\prime)} \rightarrow \bm{k} - \bm{\bar{K}}^{(\prime)}$ in Eq.~\eqref{eq:psiexp}, where $\bar{K}^{(\prime)}$ are the $K$ and $K'$ projections into the zigzag $k_y$ axis, as shown in Fig.~\ref{fig:GrapheneRibbons}(b). These $k$ projections also apply to $\bm{q}^{(\prime)}$ in $H_G(\bm{k})$ from Eq.~\eqref{eq:hq}. For simplicity, here we consider only this pristine form of $H_G(\bm{k})$. However, for narrow ribbons, symmetry-allowed corrections due to the finite size of the nanoribbons are relevant for an improved fit with \textit{ab initio} data (see Appendix \ref{app:fullk2graphene}).

The zigzag nanoribbon lattice is illustrated in Fig.~\ref{fig:GrapheneRibbons}(c). The right (left) edge ($x=\pm W_0/2$) is composed only of atoms from the $A$ ($B$) sublattice. The length $W_0 = (\frac{3}{2}N_D-1)a/\sqrt{3}$, where $N_D$ is the number of dimers. The next line of atoms, removed to form the ribbon, would have been located in $x = \pm W/2$, which defines the effective length $W = W_0 + 2a/\sqrt{3}$. Since the $p_z$ orbitals are highly localized at each carbon atom, the absence of $B$ atoms at the left edge $x = -W/2$ yields $\varphi_{B^{(\prime)}}(\bm{R}_-) \approx 0$, where $\bm{R}_\pm = (\pm W/2, y)$. Imposing $\psi(\bm{R}_-) =0$ in Eq.~\eqref{eq:psiexp}, we get $f_A(\bm{R}_-)e^{-i\bm{\bar{K}}\cdot\bm{R}_-}\varphi_A(\bm{R}_-) + f_{A'}(\bm{R}_-)e^{i\bm{\bar{K}}'\cdot\bm{R}_-}\varphi_{A'}(\bm{R}_-) = 0$. Moreover, the phase factors [see Fig.~\ref{fig:GrapheneRibbons}(a)] and orbitals cancel out, \textit{i.e.}, $e^{-i\bm{\bar{K}}\cdot\bm{R}_-}\varphi_A(\bm{R}_-) = e^{i\bm{\bar{K}}'\cdot\bm{R}_-}\varphi_{A'}(\bm{R}_-)$. Similar considerations follow for the $B$ sublattice on the right edge, $\bm{r} = \bm{R}_+$. Due to translational invariance along $y$, we can simplify $f_\mu(\bm{r}) \rightarrow e^{i k_y y}f_\mu(x)$. Therefore, the boundary conditions for the zigzag nanoribbon envelope functions are
\begin{align}
 \nonumber
 f_A\Big(-\frac{W}{2}\Big) &=  -f_{A'}\Big(-\frac{W}{2}\Big),
 \\
 f_B\Big(+\frac{W}{2}\Big) &= -f_{B'}\Big(+\frac{W}{2}\Big).
 \label{eq:BFzigzag4}
\end{align}

In the original discussion by BF \cite{BreyFertig2006}, they consider only a single valley ($K$ or $K'$) on the $\psi(\bm{r})$ expansion. Indeed, near the $K$ valley $f_{\mu'}(\bm{r}) \approx 0$ (with $\mu=A,B$). Neglecting these contributions in Eq.~\eqref{eq:BFzigzag4}, one immediately recovers their well known result $f_A(-W/2) = f_B(+W/2) = 0$. Similarly, for the $K'$ valley, one gets $f_{A'}(-W/2) = f_{B'}(+W/2) = 0$. Therefore, Eq.~\eqref{eq:BFzigzag4} is a generalization of the original BF boundary condition for the zigzag nanoribbon. 

\subsubsection{Armchair nanoribbons}

For the armchair nanoribbons, confinement along $y$ projects both $K$ and $K'$ valleys into $\bar{\Gamma}$ on the $k_x$ armchair axis [Fig.~\ref{fig:GrapheneRibbons}(b)]. Therefore, the $\psi(\bm{r})$ expansion in Eq.~\eqref{eq:psiexp} is defined by $\bm{q} = \bm{q}' \rightarrow \bm{k}$. As in the zigzag case above, these projections also apply to $H_G(\bm{k})$ in Eq.~\eqref{eq:hq}.

The armchair lattice is shown in Fig.~\ref{fig:GrapheneRibbons}(d). Confinement along $y$ defines the length $L_0 = (N_A - 1)a/2$, where $N_A$ is the number of atoms along the ribbon. The line of atoms removed to form the ribbon defines the effective length $L = L_0 + a$. Differently from the previous case, here both edge terminations contain atoms from the $A$ and $B$ sublattices. Consequently, all orbitals $\varphi_{\mu}(x,\pm L/2) \neq 0$. In this case the system has translational invariance along $y$, thus $f_\mu(\bm{r}) \rightarrow e^{i k_x x}f_\mu(y)$. Imposing $\psi(x, \pm L/2) = 0$ for all $x$ in Eq.~\eqref{eq:psiexp}, we get
\begin{align}
    f_{\mu} \Big(\pm \frac{L}{2}\Big) = -e^{\pm i\theta} f'_\mu \Big(\pm \frac{L}{2}\Big),
    \label{eq:BFarmchair}
\end{align}
where $e^{\pm i \theta} = \varphi_{\mu'}(x,\pm L/2)/\varphi_{\mu}(x,\pm L/2)$ is the  phase difference between the $K$ and $K'$ solutions at the edges. On the BF approach \cite{BreyFertig2006}, they consider the bulk Bloch phase difference, yielding $\theta \rightarrow \theta_{BF} = \Delta K \cdot L = (N_A+1)2\pi/3$. However, since confinement breaks the Bloch periodicity along $y$, deviations from $\theta_{BF}$ should be expected for narrow ribbons. Therefore, hereafter we consider Eq.~\eqref{eq:BFarmchair} with an arbitrary $\theta$ as a generalization of the original BF boundary condition.

\subsection{Equivalence between Brey and Fertig and Berry and Mondragon}\label{sec:BFequalBM}

The Brey-Fertig boundary condition discussed above, can be equally understood via the Berry-Mondragon formalism summarized in Eq.~\eqref{eq:BM}. To verify this equivalence, let us compare the BF and BM boundary conditions for the zigzag and armchair nanoribbons. Within the BM approach, the matrix $U_k$ multiplies the momentum $k$ along the confinement direction, while $U_c$ must satisfy the symmetry constraints presented previously. For the zigzag nanoribbons, these are 
\begin{align}
\nonumber
U_{k}^Z &= 
 \begin{pmatrix}
    \sigma_x & 0 \\
    0 & \sigma_x
 \end{pmatrix},
\\ 
U_c^Z(\eta) &= 
  \begin{pmatrix}
    \sigma_y\eta & \sigma_y(1-\eta) \\ 
    \sigma_y(1-\eta) & \sigma_y\eta
  \end{pmatrix},
  \label{eq:UkUcZig}
\end{align}
where $\sigma_{\nu}$ (with $\nu=0,x,y,z$)  are Pauli matrices acting on the sublattice $A$/$B$ subspace, and $\sigma_0$ is the $2\times2$ identity matrix. The parameter $\eta$ is restricted to $\eta = \{0, 1\}$, which identifies the particular boundary condition, as it is discussed below. In turn, for the armchair confinement,
\begin{align}
U_{k}^A &= 
\nonumber
 \begin{pmatrix}
    \sigma_y & 0 \\
    0 & -\sigma_y
 \end{pmatrix},
\\
U_c^A(\theta) &= 
  \begin{pmatrix}
    \sigma_y\cot\theta & \sigma_y\csc\theta \\
    \sigma_y\csc\theta & \sigma_y\cot\theta
  \end{pmatrix}.
  \label{eq:UkUcArm}
\end{align}

\textit{Armchair nanoribbons}. Substituting $U_k^A$ and $U_c^A(\theta)$ from Eq.~\eqref{eq:UkUcArm} into Eq.~\eqref{eq:BM}, one obtains Eq.~\eqref{eq:BFarmchair} after straightforward manipulations. This establishes the equivalence between the BF and BM approaches. Interestingly, these are two drastically distinct approaches for the boundary condition. On the BF approach, one uses the atomistic terminations of the lattice to motivate the boundary condition. On the other hand, the BM approach is based solely on the symmetries of the lattice.

\textit{Zigzag nanoribbons}. First, for $\eta = 0$, replacing $U_k^Z$ and $U_c^Z(0)$ from Eq.~\eqref{eq:UkUcZig} into Eq.~\eqref{eq:BM}, we reproduce our boundary conditions shown in Eq.~\eqref{eq:BFzigzag4}, plus an additional pair of equations $f_B(\bm{R}_-) = f_{B'}(\bm{R}_-)$, and $f_A(\bm{R}_+) = f_{A'}(\bm{R}_+)$, which are trivially satisfied since, $\varphi_{A^{(\prime)}}(\bm{R}_+) \approx 0$ and $\varphi_{B^{(\prime)}}(\bm{R}_-) \approx 0$. Second, for $\eta = 1$ the same procedure gives us the original BF boundary condition for the zigzag confinement. Therefore, the two possible values of $\eta = \{0, 1\}$ label our generalized boundary condition, and the original BF result. A comparison between the band structures of these two cases will be shown in the next section.

\subsection{The \texorpdfstring{$k^2$}{k²} model for graphene nanoribbons}
\label{sec:k2graphene}

To construct the $k^2$ model for the nanoribbons, we consider the symmetries of the armchair and zigzag lattices [Figs.~\ref{fig:GrapheneRibbons}(c) and \ref{fig:GrapheneRibbons}(d)]. They are both invariant under the $D_{2h}$ group. Additionally, we consider that the system is time-reversal symmetric and chiral. The matrix representations of these symmetry operations are built from the $\psi(\bm{r})$ expansion in Eq.~\eqref{eq:psiexp}. These, and the corresponding most general symmetry-allowed Hamiltonian, are shown in Appendix \ref{app:fullk2graphene}. Next, we first discuss a minimal $k^2$ model, which matches the usual $k$-linear model, but allows for simpler numerical implementations. Later, we use the full $k^2$ model to fit and compare the results with the DFT data.

\begin{figure}[b]
	\centering
	\includegraphics[width=\columnwidth]{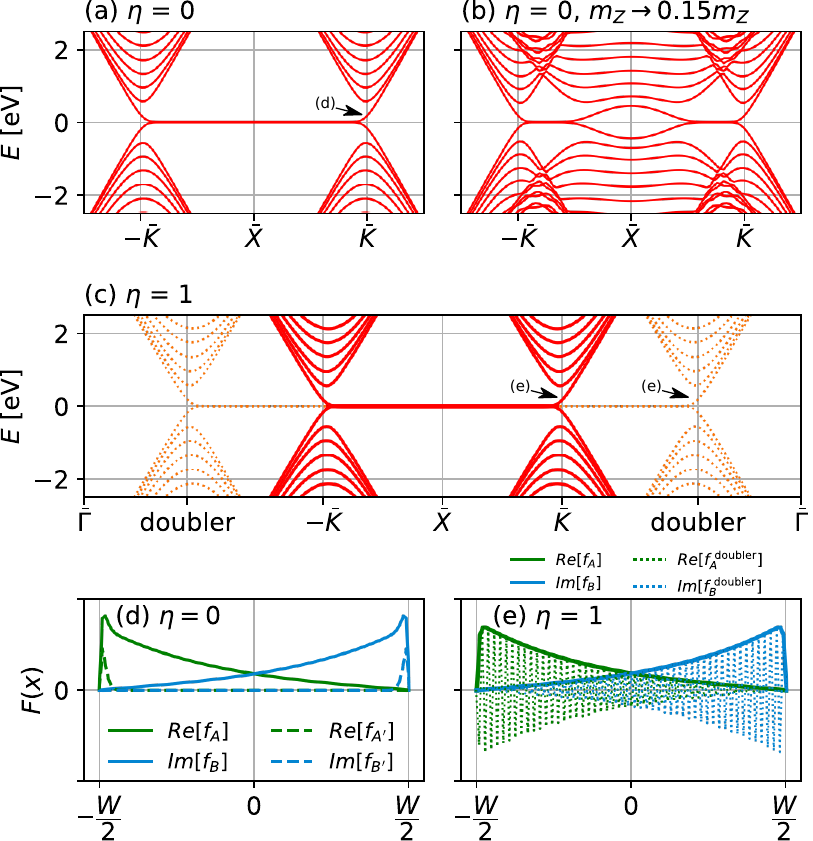}
	\caption{Zigzag bands ($N_D = 24$) from the $k^2$ model for different boundary parameters $\eta$ and Wilson's masses $m_Z$. (a) For $\eta = 0$ and finite $m_Z$, the generalized BF boundary condition returns the expected zigzag band structure. (b) For smaller $m_Z \rightarrow 0.15m_Z$ the doublers appear near $\bar{X}$ as hybridized extra cones. (c) For $\eta = 1$ (original BF) the doublers at $k_y^\text{doubler} \approx \pm[\bar{K} - 2m_Z/(\hbar v_F \delta_x^2)]$ cannot be eliminated (dotted lines). The arrows in (a) and (c) point to the state $(E,k_y)$ used to plot the envelope functions $\bm{F}(x)$.
		Near $\bar{K}$, $\bm{F}(x)$ for both (d) $\eta = 0$ and (e) $\eta = 1$ are similar and smooth (solid lines). (e) The $\bm{F}(x)$ for the doublers (dotted lines) show nonphysical phase oscillations on the scale of the numerical discretization.}
	\label{fig:zigzagbands}
\end{figure}

\subsubsection{Minimal \texorpdfstring{$k^2$}{k²} models}

A minimal model for the nanoribbons must contain only the bulk-like $k$-linear terms and the necessary $k^2$ corrections, which reads as
\begin{align}
    H_Z &= h(\bm{k}^+) \oplus h^*(\bm{k}^-)
    + \frac{m_Z}{2} U_c^Z(\eta) k_x^2,
    \label{eq:HminZ}
    \\
    H_A &= h(\bm{k}) \oplus h^*(\bm{k})
    + \frac{m_A}{2} U_c^A(\theta) k_y^2,
    \label{eq:HminA}
\end{align}
for zigzag and armchair nanoribbons, respectively. The Dirac-like term $h(\bm{k})$ is given in Eq.~\eqref{eq:hq}, and
$\bm{k}^{\pm} = (k_x, k_y \pm \bar{K})$ centers the Dirac cones into the $K$ and $K'$ projections [Fig.~\ref{fig:GrapheneRibbons}(b)]. The $k^2$ terms are defined by the Wilson's matrices $U_w^{Z(A)} \equiv U_c^{Z(A)}$ and masses $m_{Z(A)}$. Since the $k^2$ terms allow for a trivial boundary condition, \textit{i.e.,} $\psi = 0$ at the edges, it can be numerically implemented via simple finite-differences schemes \cite{Bruno2017FDP}. Hereafter, all results were obtained discretizing the coordinates with $\sim 100$ points, and the Wilson's masses are chosen on the mid-range of Eq.~\eqref{eq:mass}.

For zigzag nanoribbons, Fig.~\ref{fig:zigzagbands} compares the generalized ($\eta = 0$) and original ($\eta = 1$) BF conditions. Indeed, the ideal results are given by $\eta = 0$ in Fig.~\ref{fig:zigzagbands}(a). Here, the absence of fermion doublers is due to their hybridization near $\bar{X}$ [Fig.~\ref{fig:zigzagbands}(b)], which drives the doublers towards high energies as $m_Z$ increases. On the other hand, for $\eta = 1$ the Hamiltonian $H_Z$ splits into uncoupled $\bar{K}$ and $\bar{K}'$ blocks, each showing an independent doubler [Fig.~\ref{fig:zigzagbands}(c)]. Since there is no hybridization, the doublers will always occur at $k_y^\text{doubler} \approx \pm[\bar{K} - 2m_Z/(\hbar v_F \delta_x^2)]$. Not even staggered-lattice implementations are able to fully eliminate the doublers in this case \cite{Beenakker2008FiniteDiff, AlexisCaio2012}. Notice that in the analytical $k$-linear limit $\delta_x \rightarrow 0$ and the doublers are eliminated as $k_y^\text{doubler} \rightarrow \pm\infty$, thus justifying the use of the original BF boundary conditions in analytical calculations. For $k_y \approx \bar{K}$, the envelope functions $\bm{F}(\bm{r})$ for $\eta = \{0, 1\}$ match well, as shown in Fig.~\ref{fig:zigzagbands}(d)-(e). For $\eta=0$, a small $\bar{K}-\bar{K}'$ hybridization occurs, which is absent for $\eta=1$ due to the block form of $H_Z$. In contrast, the $\bm{F}(\bm{r})$ of the doublers show nonphysical oscillations with a period given by the numerical step size $\delta_x$ [dotted lines in Fig.~\ref{fig:zigzagbands}(e)].

\begin{figure}[b]
    \centering
    \includegraphics[width=\columnwidth]{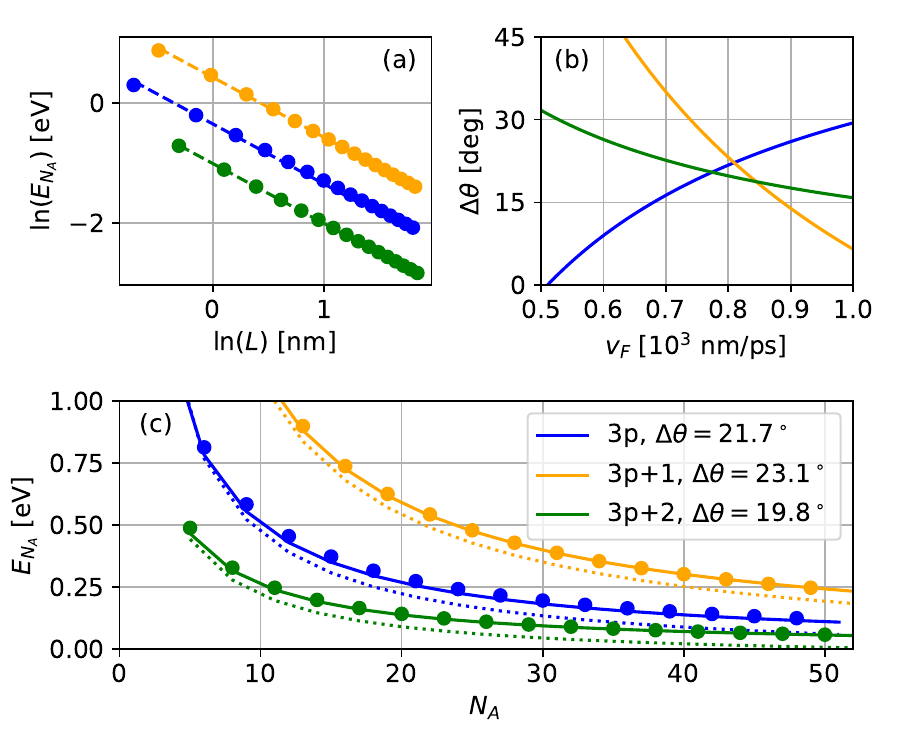}
    \caption{(a) The DFT gaps (circles) as a function of $L$ (log scale) fall into straight lines on a wide range $3 \leq N_A \leq 50$. The dashed lines are fits used to obtain the linear coefficients that yields the constraints between (b) the Fermi velocity and the correction $\Delta\theta$ to the BF boundary conditions. (c) Comparison between the gaps in DFT (circles), $k$-linear (dotted lines) and $k^2$ models (solid lines). The dotted match the solid ones, but are shifted for clarity.}
    \label{fig:armchairgap}
\end{figure}

For the armchair nanoribbons, the boundary condition is set by $\theta$. The original BF $\theta \rightarrow \theta_{BF} = (N_A+1)2\pi/3$ gives a qualitatively correct picture for the armchair gap $\propto 1/L$. However, it predicts that the bands for $N_A = 3p$ and $3p+1$ (for integer $p$) are degenerated, while for $N_A=3p+2$ it is identically gapless \cite{Son2006DFTgap}. However, as discussed above, the confinement breaks Bloch periodicity and a deviation from $\theta_{BF}$ is expected. Therefore, in Fig.~\ref{fig:armchairgap} we consider $\theta = \theta_{BF} + \Delta\theta$. Within the $k$-linear model, the armchair gaps are
\begin{align}
    E_{3p} &\approx \dfrac{\hbar v_F}{L}\Big[\dfrac{2\pi}{3} - 2\Delta\theta\Big],
    \\
    E_{3p+1} &\approx \dfrac{\hbar v_F}{L}\Big[\dfrac{2\pi}{3} + 2\Delta\theta\Big],
    \\
    E_{3p+2} &\approx \dfrac{\hbar v_F}{L}\Big[2\Delta\theta\Big].
\end{align}
Indeed, the DFT data for the gaps obey these expressions, \textit{i.e.}, $E_{N_A} \propto 1/L$, as shown in log-log scale in Fig.~\ref{fig:armchairgap}(a). The linear coefficient of these lines gives us constraints between $v_F$ and $\Delta\theta$, which we use to establish the correction $\Delta\theta$ to the BF boundary condition shown in Fig.~\ref{fig:armchairgap}(b). Considering $v_F = 0.8\times 10^3$~nm/ps (see fits in the next section), we solve the $k$-linear and $k^2$ models with the corresponding $\Delta\theta \approx 20^\circ$, and both match well the DFT gaps in Fig.~\ref{fig:armchairgap}(c). Here, we have used a constant $v_F$ for all $N_A$ for simplicity. However, $v_F$ may change as a function of $N_A$ due to finite-size effects. Consequently, $\Delta\theta$ must also be $N_A$ dependent, while obeying the constraints from Figs.~\ref{fig:armchairgap}(a) and \ref{fig:armchairgap}(b).

The comparisons above show that our $k^2$ models [Eqs.~\eqref{eq:HminZ} and \eqref{eq:HminA}] generalize the usual graphene Dirac model and the BF boundary conditions. On the zigzag case, the $\eta = 0$ model couples the $\bar{K}$ and $\bar{K}'$ valleys, such that the edge-state branch is restricted to its correct interval in Fig.~\ref{fig:zigzagbands}(a), while in the usual BF case they are uncoupled, yielding edge-state branches that extend toward  $k_y \rightarrow \pm \infty$. Indeed, both models would match identically if the valley projections are driven far apart in Eq.~\eqref{eq:HminZ}. For the armchair case, the BF model is recovered for $\Delta\theta \rightarrow 0$, yielding the zero gap for the $N_A = 3p+2$ metallic case, and degenerate $3p$ and $3p+1$ gaps. 

\subsubsection{Full \texorpdfstring{$k^2$}{k²} model: Fitting the DFT data}

\begin{figure}[b]
    \centering
    \includegraphics[width=\columnwidth]{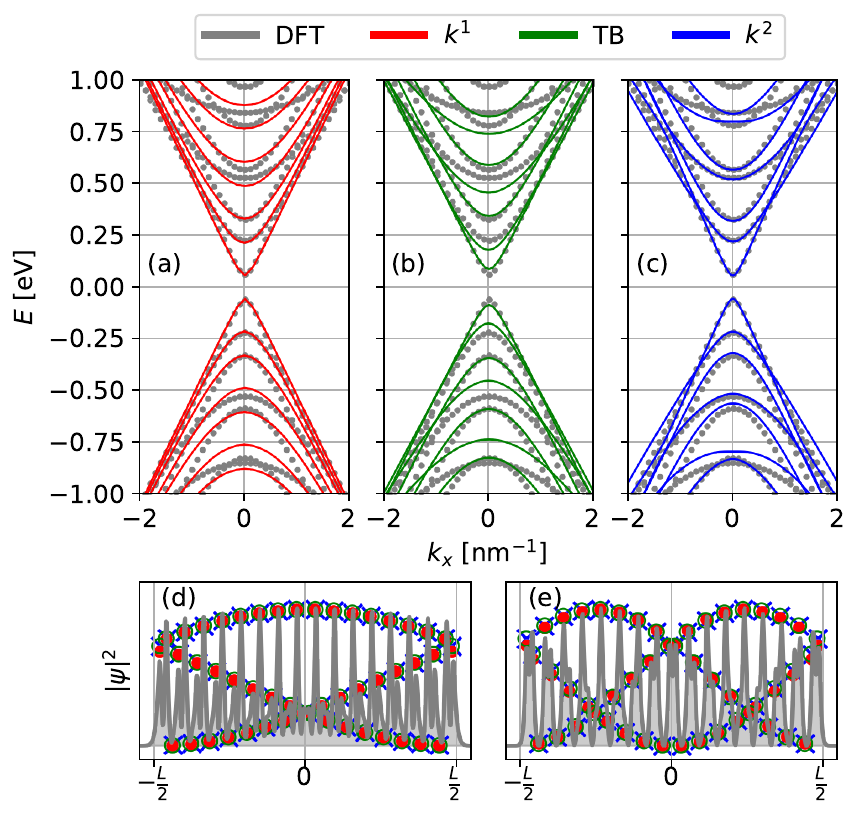}
    \caption{Comparison between the DFT band structures and (a) the $k$-linear, (b) tight-binding, and (c) $k^2$ models for $N_A = 48$. At low energy all models agree, while for $|E| \gtrsim 0.5$~eV parabolic corrections become relevant and broken chirality starts to develop. The envelope densities (colored symbols) match well the DFT data for (d) the first and (e) second conduction bands at $k_x=0$.}
    \label{fig:armchair48}
\end{figure}

While the minimal $k^2$ model above provides a sufficient approach to regularize the Dirac models on a lattice, it is also insightful to investigate the full $k^2$ model in comparison with DFT results. The derivation of the most general symmetry-allowed Hamiltonian for graphene up to $k^2$ is shown in Appendix \ref{app:fullk2graphene}. In a general compact notation, it reads as
\begin{multline}
    H = 
    h(\bm{k}^+) \oplus h^*(\bm{k}^-)
    \\
    +\Big[\dfrac{m_{A1}}{2}U_{0y}+\dfrac{m_{A2}}{2}U_{xy}\Big]k_y^2
    +\Big[\dfrac{m_{Z1}}{2}U_{0y}+\dfrac{m_{Z2}}{2}U_{xy}\Big]k_x^2
    \\
    +\Delta U_{xy}
    +\hbar\mu U_{xx}k_x
    +m_{xy}U_{zx}k_x k_y.
    \label{eq:Hfullk2}
\end{multline}
The matrices $U_{ij} = \tau_i\otimes\sigma_j$ are set in terms of the Pauli matrices $\tau_\nu$ (with $\nu=0,x,y,z$) acting on the $K$/$K'$ valley subspace, and $\sigma_\nu$ acting on the sublattices $A$/$B$. The first line in Eq.~\eqref{eq:Hfullk2} contains the Dirac-like terms. For armchair $\bm{k}^\pm = \bm{k}$, while for zigzag it sets the valley projections as discussed in the previous section. The second line shows the most general form of the $k^2$ terms. For armchair ribbons, the BM boundary conditions constrain $m_{A1} = m_A \cot\theta$ and $m_{A2} = m_A \csc\theta$. Similarly, for zigzag $m_{Z1} = m_Z \eta$ and $m_{Z2} = m_Z (1-\eta)$. The third line shows the extra terms that allow for a fine tuning of the band structure.  The $\Delta$ term couples the projected cones from $K$ and $K'$ valleys, the velocity $\mu$ couples the dispersions at finite $\bm{k}$, and $m_{xy}$ is a trigonal correction for the masses. To illustrate the results from the full $k^2$ model, we have considered a medium-sized armchair nanoribbon with $N_A = 48$ (type $3p$), as shown in Fig.~\ref{fig:armchair48}. The parameters used to obtain the figures, and equivalent results for $N_A = 49$ ($3p+1$), $N_A = 50$ ($3p+2$), and for the zigzag case are shown in Appendix \ref{app:fullk2graphene}.

In Fig.~\ref{fig:armchair48}(a)-(c) we compare the DFT band structure with the $k$-linear BF model, tight-binding (implemented with the Kwant code \cite{kwant}), and our full $k^2$ model, respectively. As expected, at low energies all models agree reasonably well with the DFT data. However, for $|E| \gtrsim 0.5$~eV discrepancies are visible in all cases. The DFT data show two sets of quantized cone dispersions with different parabolicities. This is not captured by the $k$-linear model. The tight binding model captures these features, but the band edges are shifted. The $k^2$ model provides a better fit up to $|E| \sim 0.75$~eV. 

The densities of the first and second conduction subbands are shown in Figs.~\ref{fig:armchair48}(c) and \ref{fig:armchair48}(d). The DFT data show peaks at atomic positions. For the models, the envelope functions are extracted from Eq.~\eqref{eq:psiexp}, $|\psi(y)|^2 \propto |f_\mu(y) e^{i\theta y/L} + f_{\mu'}(y) e^{-i\theta y/L}|^2$, where the phase factors arise from the $\varphi_{\mu^{(\prime)}}(\bm{r})$ phases in Fig.~\ref{fig:GrapheneRibbons}(a). These are highly oscillating envelopes, thus in Figs.~\ref{fig:armchair48}(c) and \ref{fig:armchair48}(d) we plot them only at the atomic positions, showing an excellent agreement with the DFT data. These also agree with the tight binding densities \cite{Wakabayashi2012Density} for the low-energy subbands.

\section{Spinful case: topological crystalline insulators}

\begin{figure}[b]
	\centering
	\includegraphics[width=\columnwidth]{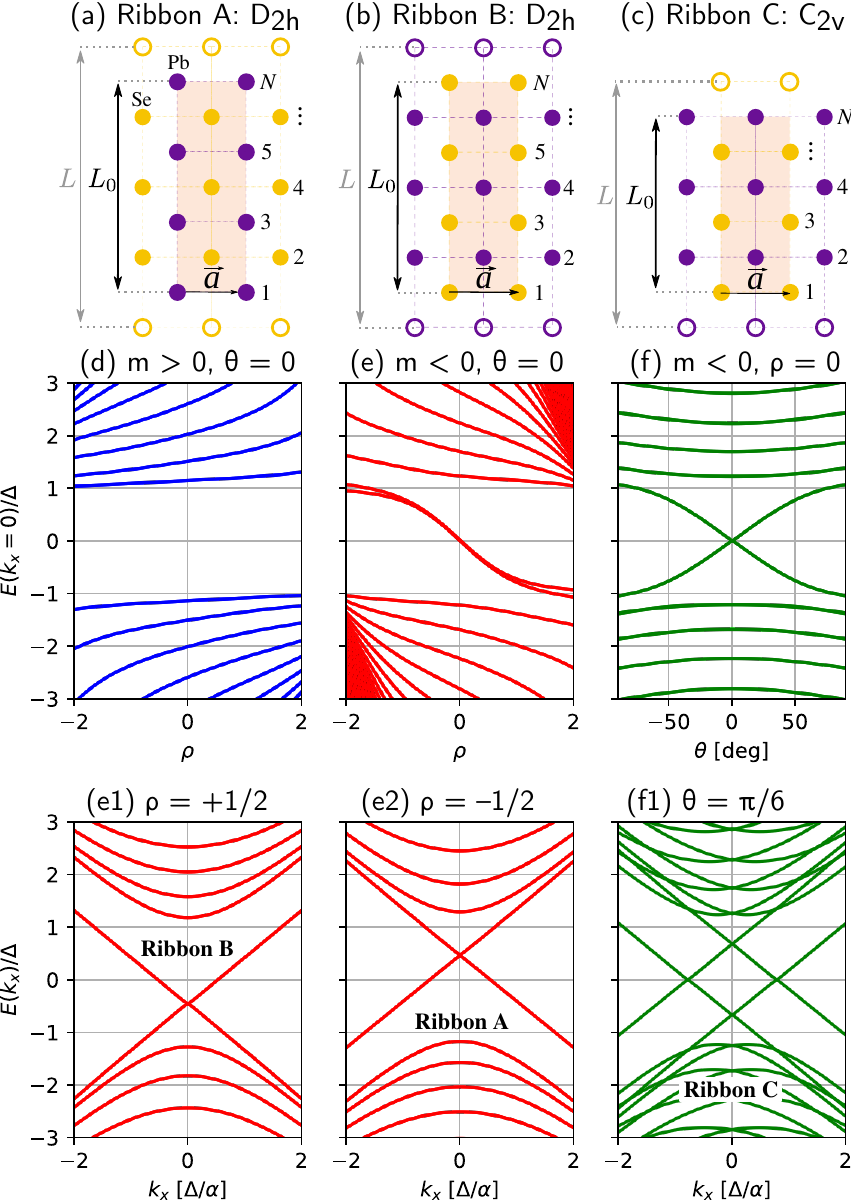}
	\caption{Lattices of PbSe ribbons of types (a) A, (b) B, and (c) C, as defined in Ref.~\cite{Araujo2016Nonsymm}.
	(d)-(f) PbSe band edges at $k_x=0$ varying the boundary condition parameters $\rho$ and $\theta$. (d) For $m>0$, the system is trivial, there are no states within the gap $|E/\Delta| < 1$. (e) For $m<0$, a pair of degenerate topological Dirac crossings appear, and its crossing point at $k_x=0$ is controlled by $\rho$: (e1) for $\rho>0$ the crossing is down-shifted, and (e2) for $\rho<0$ it shifts up in energy. (f) A finite $\theta$ or $\Delta_C$ splits the Dirac crossings as shown in (f1).}
	\label{fig:pbse}
\end{figure}

Graphene has a very weak spin-orbit coupling \cite{Graphene1,Graphene2,CastroNeto2009Review}. Therefore, to illustrate our results on a spinful system, let us instead consider a monolayer of PbSe, which is a topological crystalline insulator (TCI) \cite{PbSe1,PbSe2,PbSe3}. The effective model for this material was derived in Ref.~\cite{Araujo2016Nonsymm} up to $k^2$. Hereafter, we follow the notation from this reference. For simplicity, we restrict the discussion to the PbSe nanoribbons of types A, B and C (Fig.~\ref{fig:pbse}), while generalizations for ribbons D and E (defined in Ref.~\cite{Araujo2016Nonsymm} and not shown here) are straightforward.

The lattices from ribbons A and B are invariant under the point group $D_{2h}$, while for ribbon C the symmetry is reduced to $C_{2v}$. Therefore, the sole difference between ribbons A and B is their atomic terminations [see  Fig.~\ref{fig:pbse}], which shall reflect on their boundary conditions. For ribbon C, one edge is equivalent to that of ribbon A, and the other is of the B type. Consequently, due to the reduced symmetry, ribbon C admits extra terms in the Hamiltonian.

The model of PbSe monolayers \cite{Araujo2016Nonsymm} around the X point of the Brillouin zone is defined on the basis functions $\{ \varphi_{xz,\uparrow}(\bm{r}), \varphi_{xz,\downarrow}(\bm{r}), \varphi_{x,\uparrow}(\bm{r}), \varphi_{x,\downarrow}(\bm{r}) \}$, where the $xz$ and $x$ indices refer to the symmetries of the orbitals, and $\uparrow, \downarrow$ label the spin states along $z$. Thus, similarly to Eq.~\eqref{eq:psiexp}, the wave-function $\psi(\bm{r})$ expansion is given by these basis functions multiplied by the $k$-phase $e^{i(\bm{k}-\bm{X})\cdot\bm{r}}$ and the envelope spinor $\bm{F}(x,y) = e^{i k_x x}[f_{xz,\uparrow}(y), f_{xz,\downarrow}(y), f_{x,\uparrow}(y), f_{x,\downarrow}(y)]^T$. Here we already assume a plane-wave along $x$, since the confinement in along $y$ in Figs.~\ref{fig:pbse}(a)-(c).

Considering an isotropic limit for simplicity, the \textit{minimal} $k^2$ effective model for ribbons A, B and C, confined along $y$ and extended along $x$, is
\begin{multline}
	H = \Delta U_{z0} + \alpha(U_{xx}k_y - U_{xy}k_x) +\frac{m}{2} U_c(\rho, \theta) k_y^2
	+ \Delta_C U_{yx},
\end{multline}
\vspace*{-0.5cm}
\begin{multline}
	U_c(\rho, \theta) = \cos\theta\Big[\cosh(\rho) U_{z0} + \sinh(\rho) U_{00} \Big] + \sin(\theta) U_{yx}.
\end{multline}
The matrices $U_{ij} = \tau_i \otimes \sigma_j$ are set by Pauli matrices acting on the orbital ($\tau_\nu$) and spin ($\sigma_\nu$) subspaces, $\Delta$ is the gap at $\bm{k} = 0$, $\alpha$ defines the Fermi velocity, and $m$ is Wilson's mass. The coupling $\Delta_C$ is only allowed for ribbon C. Since the confinement is along $y$ the kinetic matrix $U_k = U_{xx}$, yielding $\det[U_k \pm i U_c(\rho,\theta)] = 0$, as expected. The boundary conditions are defined by the continuum parameters $\rho$ and $\theta$. For ribbons A and B, $\theta \equiv 0$.

The effects of the boundary condition parameter $\rho$ and $\theta$ on the band structure are shown in Figs.~\ref{fig:pbse}(d)-\ref{fig:pbse}(f). Here, we consider $\Delta$ and $\alpha/\Delta$ as the energy and distance units. For $m>0$ the system is trivial, thus there are no states within the gap $|E/\Delta| < 1$ in Fig.~\ref{fig:pbse}(d). For $m<0$ the system becomes topologically non-trivial with a mirror Chern number $n_M = -2$ \cite{Araujo2016Nonsymm}, yielding two Dirac cones (degenerate for $\theta=0$). In this case, the states seen within the gap in Fig. \ref{fig:pbse}(e) refer to the crossing point of the Dirac dispersion seen in Figs.~\ref{fig:pbse}(e1) and \ref{fig:pbse}(e2). The system is chiral for $\rho = 0$, while for finite $\rho$ the broken chirality is a consequence of the distinct atomic terminations of ribbons A and B \cite{Araujo2016Nonsymm}. For ribbon C we can consider $\rho = 0$ for simplicity and allow $\theta$ to vary. This is shown in Fig.~\ref{fig:pbse}(f). In this case, both $\theta \neq 0$ or $\Delta_C\neq 0$ break the degeneracies between the Dirac crossings, as seen in Fig.~\ref{fig:pbse}(f1). These results are equivalent to those from Ref.~\cite{Araujo2016Nonsymm}, where a BF-type boundary condition was proposed.

Complementarily, within the $k$-linear model ($m=0$), the BM approach for the boundary conditions $[U_k \pm i U_c(\rho,\theta)]\cdot\bm{F}(x,\pm L/2) = 0$ yields,
\begin{align}
	f_{x,+\sigma}\Big(\pm \frac{L}{2}\Big) &= \mp \dfrac{i e^{\rho} \cos\theta}{1 \pm \sin\theta}f_{xz,-\sigma}\Big(\pm \frac{L}{2}\Big).
\end{align}
Interestingly, this boundary condition implies a spin-orbital admixture, as it couples opposite spins $\pm\sigma$ and orbitals $x/xz$ \cite{noteBCprev}. In Ref.~\cite{Ferreira2013Magnetically}, this type of constraint leads to a spin texture across the ribbon.

\section{Conclusions}

We have shown how the $k^2$ Wilson corrections not only regularize the Dirac-like models on a lattice, but are directly related to the boundary conditions of finite systems. Considering the symmetries of the finite size system (\textit{e.g.,} nanoribbons), the choice of Wilson's corrections are not arbitrary. Indeed, we show that the symmetry-allowed $k^2$ terms are equivalent to the non-trivial boundary conditions from Berry and Mondragon \cite{BerryMondragon1987}, thus providing a recipe to regularize the Dirac model by including the $k^2$ term compatible with the desired boundary condition. This hidden connection between Wilson's $k^2$ term and the boundary conditions were taken as a conjecture in Ref.~\cite{Bruno2017FDP} to propose a simple method to eliminate the fermion doubling problem. Here, our systematic derivation now proves this conjecture.

Applying this methodology for graphene, we have found a generalization of the Brey-Fertig boundary conditions \cite{BreyFertig2006}. For the zigzag nanoribbons, the $K$--$K'$ coupling induced by the boundary condition restricts the edge-state bands to lie within these valleys. More interestingly, for the armchair case, it introduces $\Delta\theta$ as a deviation from the bulk Bloch phases. Particularly, for the ``metallic'' armchair case a finite $\Delta\theta$ eliminates the nonphysical gapless band structure \cite{Son2006DFTgap}. Additionally, for the spinful systems (\textit{e.g.,} PbSe TCI) our approach allows for simple derivation of the spinful boundary conditions.

\section{Acknowledgements}

We thank M. Novaes and L. Chico for useful discussions. A.L.A., R.P.M, R.G.F.D., and G.J.F. acknowledge financial support from the Brazilian funding agencies
CNPq, CAPES, and FAPEMIG, and D.V. acknowledges funding from the Dutch national
science organization (NWO) VIDI grant 680-47-53. 

A.L.A. and R.P.M. have contributed equally to this work.


\appendix

\section{Rigorous proofs of the hard-wall boundary conditions}
\label{app:rigorous}

Section \S\ref{sec:hardwall} of the main text presents an illustrative picture on how the boundary conditions change from the $k$-linear to the $k^2$ model. Here, we provide more rigorous proofs of the equivalence of these boundary conditions for the general case of $N\times N$ (for even $N$) Hamiltonians with Dirac-like spectrum.

\subsection{Constraints on $U_c$ and $U_w$}

In the linear model, to define a hard wall, $U_c$ needs to open a gap at zero energy in the outside region. Consequently, the Hamiltonian $(\hbar v_F U_k k + V_0 U_c)$ should have no propagating modes at zero energy, meaning the determinant cannot have any zeros for real $k$.
Using that $U_k$ is nonsingular, this is equivalent to $U_k^{-1} U_c$ having no real eigenvalues, which shows that $U_c$ cannot be singular.
Demanding that $U_c$ anticommutes with $U_k$ is sufficient to fulfill this condition, as this ensures that $U_k^{-1} U_c$ is skew-Hermitian with purely imaginary eigenvalues.

Similarly, in the quadratic case, the Wilson's mass term cannot introduce any new modes, \textit{i.e.,} $(\hbar v_F U_k k + m U_w k^2)$ should have no propagating modes at zero energy beyond the $N$ original $k=0$ modes of the linear model.
Furthermore, the Wilson's mass term should ensure that in the discretized model [$k \to \sin(\delta_x k)/\delta_x$] has no fermion doubling at $k = \pi / \delta_x$, requiring non-singular $U_w$.
Dividing the Hamiltonian by $k$ we get $(\hbar v_F U_k + m U_w k)$ whose determinant should not vanish for any real $k$.
Using that $U_k$ and $U_w$ are non-singular, this is equivalent to $U_k^{-1} U_w$ having no real eigenvalues.

Hence, we showed that $U_c$ and $U_w$ obey the same constraint, and demanding anticommutation with $U_k$ is sufficient to fulfill it.

\subsection{Boundary conditions for the $k$-linear model}
\label{app:k_linear}

Consider the general model Hamiltonian from Eq.~\eqref{eq:genH} in the $k$-linear case ($m=0$), \textit{i.e.}, $H = \hbar v_F U_k k + U_c V(x)$, and its corresponding Schrödinger equation $H F(x) = E F(x)$. Let us assume that $V(x) = V_0 \Theta(x)$ defines a single wall at $x=0$, opening a gap $|2V_0|$ for $x>0$. On the outer region, the hard-wall limit $(V_0\rightarrow \infty$) allow us to neglect $E \ll V_0$, and considering the ansatz $F(x>0) \sim F_0 e^{-x/\lambda}$ ($\Re\lambda >0$), we get
\begin{align}
	\Big(\dfrac{i\hbar v_F}{V_0\lambda} U_k + U_c\Big) F_0 = 0.
	\label{eq:BClambda}
\end{align}
Non-trivial solutions will only exist if the determinant of the matrix in parentheses vanishes, which can be cast as
\begin{align}
	\det \Big( U_k^{-1}U_c - \omega\Big) = 0,
	\label{eq:UkcEig}
\end{align}
since $U_k$ is non-singular. Here, $\omega = -i\hbar v_F/V_0\lambda$ plays the role of the eigenvalues of $U_k^{-1}U_c$. The restriction $\Re\lambda >0$, imposed by the evanescent solution for $x>0$, implies $\Im\omega < 0$. Next, one must check that $U_k^{-1}U_c$ admits eigenvalues with $\Im\omega<0$.

First, notice that Eq.~\eqref{eq:BClambda} is formally identical to the constraint on $U_c$ derived in the previous section, requiring that there are no solutions with purely real $\omega$. Second, using that $U_k^{-1}U_c$ is a product of two Hermitian matrices, it is easy to prove that its eigenvalues come in complex-conjugate pairs, i.e., simply take the complex conjugate of Eq.~\eqref{eq:UkcEig} to find
\begin{align}
	\label{eq:conjugate}
	\nonumber
	0 &= \det\Big(U_c - \omega U_k\Big) = \det\Big(U_c - \omega U_k\Big)^*  \nonumber\\
	&= \det\Big[(U_c - \omega U_k)^\dagger\Big] = \det\Big(U_c - \omega^* U_k\Big).
\end{align}
Therefore, both $\omega$ and $\omega^*$ are eigenvalues of $U_k^{-1}U_c$, and $*$ labels complex conjugation. This guarantees that there are always exactly $N/2$ solutions with $\Im\omega < 0$.

Finally, we check that all solutions of this boundary condition obey current conservation, meaning vanishing normal current for any solution of a hard-wall boundary condition.
The current operator normal to the boundary is given by
\begin{equation}
J = \frac{\partial H}{\partial (\hbar k)} = v_F U_k,
\end{equation}
and the expectation value of the current is
\begin{align}
F_0^{\dag} J F_0 & = v_F F_0^{\dag} U_k F_0 \nonumber\\
&= \omega^{-1} v_F F_0^{\dag} U_k (U_k^{-1} U_c) F_0  \nonumber\\
&= \omega^{-1} v_F F_0^{\dag} U_c F_0,
\end{align}
where we used that $F_0$ is a solution of the boundary condition.
The left-hand side is the expectation value of a Hermitian operator and is real, while the right hand side is a nonzero, not real number times another real expectation value. This is only possible if $F_0^{\dag} U_k F_0 = F_0^{\dag} U_c F_0 = 0$. This shows that the boundary condition is current conserving.

\subsection{Boundary conditions for the $k^2$ model}
\label{app:k_square}

For the full model Hamiltonian from Eq.~\eqref{eq:genH}, we expect that the hard-wall boundary condition becomes $F(0) = 0$ at the hard-wall set at $x=0$. This is only possible if the bulk solutions in the inner region [$x<0$, thus $V(x)=0$] are linearly dependent, thus allowing for a linear combination that vanishes at the boundary.

To verify this condition, consider a plane-wave ansatz $F(x<0) \sim F_k e^{ikx}$, with $\Im k\leq 0$ to avoid divergent states for $x\rightarrow -\infty$. The Schrödinger equation for $x<0$ becomes
\begin{align}
	\Big(\hbar v_F U_k k + m U_w k^2 - E\Big)F_k = 0.
\end{align}
Nontrivial solutions require $\det[S(k,E)] = 0$, with $S(k,E) = \hbar v_F U_k k + m U_w k^2 - E$, yielding a $2N^{\rm th}$-order polynomial equation for $k$, where $N$ is the size of the matrix $S(k,E)$. For a fixed $E$, this provides $2N$ complex roots $k_n$, which are either purely real, or come in complex-conjugate pairs, as shown by a similar argument to Eq.~\eqref{eq:conjugate}. Together with the constraint on $U_w$ this guarantees that there are $N$ bulk states at a given $E$ with purely real $k_n$, and a total of $N+N/2$ roots with $\Im k_n \leq 0$. Since the nullity of $S(k,E)$ cannot be larger than $N$, this implies that the set of zero eigenvectors $F_{k_n}$ with $\Im k_n < 0$ is linearly dependent, thus validating the criteria stated above.

\subsection{Relation between the wave functions of the $k$-linear and $k^2$ models}

To further validate the equivalence between the $k$-linear and $k^2$ models, we show that their envelope functions $F(x)$ are closely related in the limit of small energies $E \approx 0$ and momentum $k \approx 0$, for which the linear spectrum dominates.

First, consider the linear model $H$ from Eq.~\eqref{eq:Hk1}, and a plane-wave ansatz $F(x) = F_n e^{ikx}$. Since $[H, U_k] = 0$, they share a common set of eigenmodes $F_n$, which does not depend on $k$. The set $F_n$ spans the full internal Hilbert space, thus allowing for a linear combination that satisfies the boundary condition from Eq.~\eqref{eq:Hk1} (see Appendix \ref{app:k_linear}).

Second, for the $k^2$ model from Eq.~\eqref{eq:Hk2}, the boundary condition changes to $F(0) = 0$ (see Appendix \ref{app:k_square}). However, in the small-$k$ limit we can neglect the $k^2$ term, hence, the small-$k$ eigenmodes are approximately the same set $F_n$ from the $k$-linear case. Since this set is a complete basis, the boundary condition $F(0) = 0$ can only be set by the trivially zero linear combination. Therefore, this boundary condition requires the modes from higher momentum states. To obtain these, consider the small energy limit ($E\rightarrow 0$) on the Schrödinger equation for the $k^2$ model, and simplifying $k$ in the equation to eliminate the small-$k$ solutions, we get
\begin{align}
	(\hbar v_F U_k + mU_w k)F_k = 0.
\end{align}
As shown in Appendix \ref{app:k_square}, this equation is guaranteed to provide evanescent solutions that decay toward the interior of the sample, where $k_e = \hbar v_F/(m \omega)$ with $\Im k_e > 0$, and $\omega$ is a complex eigenvalue of $U_k^{-1}U_w$ with $\Im \omega > 0$. This equation is formally identical to the boundary condition of the $k$-linear model~\eqref{eq:linearBC}. Since the small-$k$ modes $F_n$ form a complete set, the addition of these $F_{k_e}$ evanescent modes makes the set linearly dependent, allowing a vanishing combination at the boundary. For small $m$, $k_e$ is large and the evanescent solutions decay quickly, while the propagating part of the solution with small $k$ satisfies the same boundary condition as the linear model.

The analysis above shows that the envelope functions of the two models match far from the boundaries. Near the boundary, the $k^2$ model has an additional localized contribution that decays quickly away from the boundary in the limit of small $m$~\cite{massRange}.

\section{Full \texorpdfstring{$k^2$}{k²} model for graphene}
\label{app:fullk2graphene}

The full monolayer graphene honeycomb lattice is invariant under the ($P6/mmm$) symmorphic space group. Eliminating the trivial Bloch translations group $T_B$, its factor point group is ($P6/mmm$)/$T_B \sim D_{6h}$. On the other hand, once the monolayer is cut to form the nanoribbons, the sixfold rotation symmetry is broken, such that the zigzag and armchair nanoribbons may transform as either the $Pmmm$ or $Pmma$ space groups, depending on their widths.

Armchair nanoribbons with odd $N_A$ transform as the symmorphic $Pmmm$ space group, while for even $N_A$, it transforms as the nonsymmorphic $Pmma$. Nevertheless, since the monolayer Dirac cones are projected into $\bar{\Gamma}$ for the armchair confinement, both have the same factor group under trivial Bloch translations, thus, ${Pmmm}/T_B \equiv {Pmma}/T_B \sim D_{2h}$. Similarly, zigzag nanoribbons with even (odd) $N_D$ belong to the $Pmmm$ ($Pmma$) space group. In this case, the $K$ and $K'$ monolayer valleys fall into the $\bar{K}$ and $\bar{K}'$ under the confinement projection. Consequently, the $Pmma$ nonsymmorphic symmetries yield an extra phase into the representation matrices. Fortunately, this phase matches that of a single Bloch translation, thus, it also follows $\text{Pmmm}/T_B \equiv \text{Pmma}/T_B \sim D_{2h}$. Therefore, hereafter it is sufficient to analyze the symmorphic lattices and the point group $D_{2h}$. This allows us to build a single model for both armchair and zigzag nanoribbons, considering a basis set that contains both $K$ and $K'$ basis functions.

The $D_{2h}$ point group can be generated by its mirror operations $M_x$, $M_y$, and $M_z$. Here, $M_x$ reflects $x \rightarrow -x$, and similarly for $M_{y}$ and $M_{z}$. Under the basis vector $\bm{r} = (x,y,z)$, the coordinates representation from the $O(3)$ group is given by the matrices
\begin{align}
    D^{\bm{r}}(M_x) &= \text{diag}(-1, +1, +1),
    \\
    D^{\bm{r}}(M_y) &= \text{diag}(+1, -1, +1),
    \\
    D^{\bm{r}}(M_z) &= \text{diag}(+1, +1, -1),
\end{align}
where $\text{diag}(\cdots)$ labels a diagonal matrix with elements given by its arguments.

To obtain the Hilbert space representation $\mathcal{H}$ we consider the basis functions $\{\varphi_A(\bm{r}), \varphi_B(\bm{r}), \varphi_{A'}(\bm{r}), \varphi_{B'}(\bm{r})\}$ shown in Fig.~\ref{fig:GrapheneRibbons}(a). These are built from $p_z$ orbitals of the carbon atoms centered at the $A$ or $B$ lattice sites and Bloch phases related to the $K$ or $K'$ valleys. Namely, the representation matrices for the $D_{2h}$ generators are
\begin{align}
    D^\mathcal{H}(M_x) &= -\tau_0 \otimes \sigma_y,
    \\
    D^\mathcal{H}(M_y) &= +\tau_x \otimes \sigma_0,
    \\
    D^\mathcal{H}(M_z) &= -\tau_0 \otimes \sigma_0,
\end{align}
where $\bm{\sigma} = (\sigma_0, \sigma_x, \sigma_y, \sigma_z)$ are Pauli matrices acting on the $A$/$B$ lattices subspace, and $\bm{\tau} = (\tau_0, \tau_x, \tau_y, \tau_z)$ acts on the $K/K'$ valley subspace.

Additionally, we consider that the system is chiral $\mathcal{C}$ and time-reversal $\mathcal{T}$ invariant. Since our graphene model is spinless, $\mathcal{T} = \mathcal{K}$ is simply the complex conjugation. The $\mathcal{C}$ symmetry labels the sublattices. Under the $\bm{r}$-representation $D^{\bm{r}}(\mathcal{T}) = D^{\bm{r}}(\mathcal{C}) = \mathds{1}$, while on $\bm{k}$ space $D^{\bm{k}}(\mathcal{T}) = -\mathds{1}$ and $D^{\bm{k}}(\mathcal{C}) = \mathds{1}$. Within the $\mathcal{H}$ representation,
\begin{align}
    D^{\mathcal{H}}(\mathcal{T}) &= \tau_x \otimes \sigma_z \mathcal{K},
    \\
    D^{\mathcal{H}}(\mathcal{C}) &= \tau_0 \otimes \sigma_z.
\end{align}

To obtain the effective Hamiltonian $H$, we consider the method of invariants \cite{winkler2003spin}. Thus, we seek the most general form of $H \equiv H(\bm{k})$ as an expansion in powers of $\bm{k} = (k_x, k_y)$ that is invariant, \textit{i.e.}, $[H, \mathcal{S}] = 0$ for all $\mathcal{S}$ symmetries above. This can be easily implemented in using the QSYMM Python's package \cite{Qsymm2018}. Splitting the resulting terms as $H = H_0 + H_A + H_Z + H_{\rm ft}$, we obtain

\begin{widetext}
\begin{align}
    H_0 &= 
        \hbar v_x
        \begin{pmatrix}
            \sigma_x & 0 \\
            0 & \sigma_x
        \end{pmatrix} k_x
        +
        \hbar v_y\Big[
        \begin{pmatrix}
            \sigma_y  & 0 \\
            0 & -\sigma_y
        \end{pmatrix} k_y
        +
        \begin{pmatrix}
            \sigma_y & 0 \\
            0 & \sigma_y
        \end{pmatrix}
        \Delta_K\Big],
    \label{eq:HZA0}
\\
    H_A &= \dfrac{m_{A1}}{2}
        \begin{pmatrix}
            \sigma_y & 0 \\
            0 & \sigma_y
        \end{pmatrix} k_y^2
        +
        \dfrac{m_{A2}}{2}
        \begin{pmatrix}
            0 & \sigma_y \\
            \sigma_y & 0
        \end{pmatrix} k_y^2
        \longrightarrow
        H_A = 
        \dfrac{m_A}{2}
        \begin{pmatrix}
            \sigma_y\cot\theta & \sigma_y\csc\theta \\
            \sigma_y\csc\theta & \sigma_y\cot\theta
        \end{pmatrix} k_y^2,
        \label{eq:HA}
\\
    H_Z &= \dfrac{m_{Z1}}{2}
        \begin{pmatrix}
            \sigma_y & 0 \\
            0 & \sigma_y
        \end{pmatrix} k_x^2
        +
        \dfrac{m_{Z2}}{2}
        \begin{pmatrix}
            0 & \sigma_y \\
            \sigma_y & 0
        \end{pmatrix} k_x^2
        \longrightarrow
        H_Z = 
        \dfrac{m_Z}{2}
        \begin{pmatrix}
            \eta\sigma_y & (1-\eta)\sigma_y \\
            (1-\eta)\sigma_y & \eta\sigma_y 
        \end{pmatrix} k_x^2,
        \label{eq:HZ}
\\
    H_{\rm ft} &= 
    \Delta
    \begin{pmatrix}
        0 & \sigma_y \\
        \sigma_y & 0
    \end{pmatrix}
    +
    \hbar \mu
    \begin{pmatrix}
        0 & \sigma_x \\
        \sigma_x & 0
    \end{pmatrix} k_x
    +
    m_{xy}
    \begin{pmatrix}
        \sigma_x & 0 \\
        0 & -\sigma_x 
        \end{pmatrix}k_x k_y.
    \label{eq:Ha1}
\end{align}
\end{widetext}
Here, $H_0$ represents a minimal Dirac-like model with anisotropic Fermi velocities $v_{x}$ and $v_{y}$, with the monolayer projected cones at $k_y = \pm \Delta_K = \pm \bar{K}$ for the zigzag confinement, and $\Delta_K = 0$ for armchair. The $H_A$ and $H_Z$ show the most general $k_x^2$ and $k_y^2$ terms on the left hand side. On the right hand side, the second form of $H_A$ and $H_Z$ are written as the $U_c \equiv U_w$ confinement or Wilson's matrices for the \textit{minimal} armchair and zigzag models, respectively. Additionally, $H_{\rm ft}$ contains fine-tuning terms that were neglected on the minimal models [Eqs.~\eqref{eq:HminZ} and \eqref{eq:HminA}].

\begin{figure*}[ht!]
    \centering
    \includegraphics[width=\textwidth]{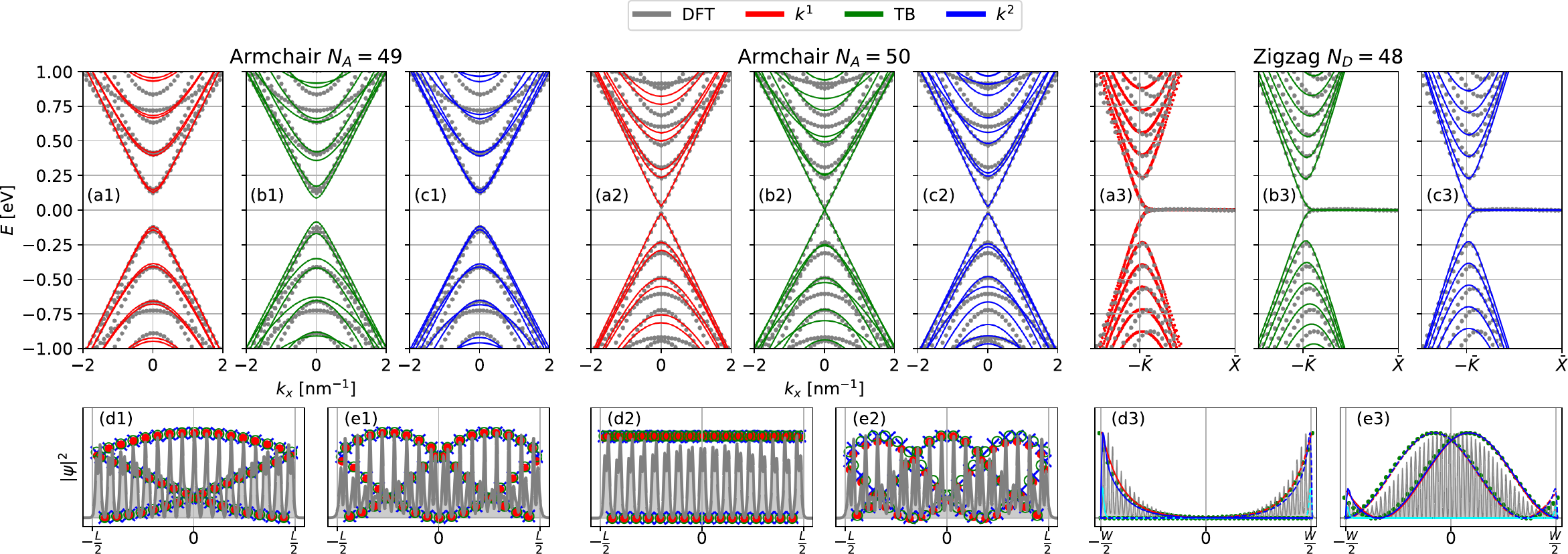}
    \caption{Comparison between the DFT band structures and (a) the $k$-linear, (b) tight-binding, and (c) $k^2$ models for armchair nanoribbons with (a1)-(c1) $N_A = 49$, (a2)-(c2) $N_A = 50$, and (a3)-(c3) zigzag nanoribbons with $N_D=48$. The envelope densities (colored symbols) match well the DFT data for (d) the first and (e) second conduction bands at $k_x=0$ in all cases.}
    \label{fig:az-combo}
\end{figure*}

We have fit the full $k^2$ model to the DFT data, trying to use a minimal set of finite parameters (see Table \ref{tab:parameters}). The results for the $N_A = 48$ armchair ribbon was shown in the main text, Fig.~\ref{fig:armchair48}. Here we show equivalent results for $N_A=49$ in Figs.~\ref{fig:az-combo}(a1)-(e1), for the metallic case $N_A=50$ in Figs.~\ref{fig:az-combo}(a2)-(e2), and for a zigzag ribbon with $N_D=48$ in Figs.~\ref{fig:az-combo}(a3)-(e3). In all cases, the agreement between the models and the DFT data at low energies is satisfactory. At higher energies, the $k^2$ model provides slightly better results. Since all models are based on low-energy expansions, the agreement with DFT must improve for wider ribbons. However, the main conclusion to be extracted from the comparison is that the $k^2$ model provides a better approach for numerical simulations in comparison to the $k$-linear model, which requires complex handling \cite{Beenakker2008FiniteDiff, AlexisCaio2012}.

\begin{table}[b]
\centering
\caption{Parameters used for the $k$-linear, $k^2$, and tight-binding models for the armchair nanoribbons.}
\label{tab:parameters}
\resizebox{\columnwidth}{!}{%
\begin{tabular}{cccccc}
\hline
\multicolumn{1}{c|}{}                                        & \multicolumn{3}{c|}{Armchair}            & \multicolumn{2}{c}{Zigzag}                   \\ \hline
\multicolumn{1}{c|}{$N_A$}                                   & 48   & 49    & \multicolumn{1}{c|}{50}   & \multicolumn{1}{c|}{$N_D$}             & 48  \\ \hline
\multicolumn{6}{c}{$k$-linear model}                                                                                                                   \\ \hline
\multicolumn{1}{c|}{$v_F$ {[}$10^3$ nm/ps{]}}                & 0.8  & 0.8   & \multicolumn{1}{c|}{0.8}  & \multicolumn{1}{c|}{$v_F$}             & 0.8 \\
\multicolumn{1}{c|}{$\Delta\theta$ {[}deg{]}}                & 21.7 & 23.1  & \multicolumn{1}{c|}{19.8} & \multicolumn{1}{c|}{-}                 & -   \\ \hline
\multicolumn{6}{c}{$k^2$ model}                                                                                                                        \\ \hline
\multicolumn{1}{c|}{$v_x = v_y$ {[}$10^3$ nm/ps{]}}          & 0.8  & 0.8   & \multicolumn{1}{c|}{0.8}  & \multicolumn{1}{c|}{$v_x = v_y$}       & 0.8 \\
\multicolumn{1}{c|}{$m_A$ {[}meV~nm$^2${]}}             & 50   & 50    & \multicolumn{1}{c|}{50}   & \multicolumn{1}{c|}{$m_Z$}             & 50  \\
\multicolumn{1}{c|}{$\Delta\theta$ {[}deg{]}}                & 21.7 & 23.1  & \multicolumn{1}{c|}{19.8} & \multicolumn{1}{c|}{$\eta$}            & 0   \\
\multicolumn{1}{c|}{$m_{Z1} = m_{Z2}$ {[}meV~nm$^2${]}} & 0    & 0     & \multicolumn{1}{c|}{0}    & \multicolumn{1}{c|}{$m_{A1} = m_{A2}$} & 0   \\
\multicolumn{1}{c|}{$\mu$ {[}$10^3$ nm/ps{]}}                & 0.1  & -0.05 & \multicolumn{1}{c|}{0}    & \multicolumn{1}{c|}{$\mu$}             & 0   \\
\multicolumn{1}{c|}{$\Delta$ {[}meV{]}}                      & 0    & 0     & \multicolumn{1}{c|}{0}    & \multicolumn{1}{c|}{$\Delta$}          & 0   \\
\multicolumn{1}{c|}{$m_{xy}$ {[}meV~nm$^2${]}}          & -50    & -50     & \multicolumn{1}{c|}{-50}    & \multicolumn{1}{c|}{$m_{xy}$}          & 0   \\ \hline
\multicolumn{6}{c}{Tight binding}                                                                                                                      \\ \hline
\multicolumn{1}{c|}{Hopping {[}eV{]}}                        & 2.4  & 2.4   & \multicolumn{1}{c|}{2.4}  & \multicolumn{1}{c|}{Hopping}           & 2.4 \\ \hline
\end{tabular}%
}
\end{table}

\bibliography{main}

\begin{thebibliography}{62}%
\makeatletter
\providecommand \@ifxundefined [1]{%
 \@ifx{#1\undefined}
}%
\providecommand \@ifnum [1]{%
 \ifnum #1\expandafter \@firstoftwo
 \else \expandafter \@secondoftwo
 \fi
}%
\providecommand \@ifx [1]{%
 \ifx #1\expandafter \@firstoftwo
 \else \expandafter \@secondoftwo
 \fi
}%
\providecommand \natexlab [1]{#1}%
\providecommand \enquote  [1]{``#1''}%
\providecommand \bibnamefont  [1]{#1}%
\providecommand \bibfnamefont [1]{#1}%
\providecommand \citenamefont [1]{#1}%
\providecommand \href@noop [0]{\@secondoftwo}%
\providecommand \href [0]{\begingroup \@sanitize@url \@href}%
\providecommand \@href[1]{\@@startlink{#1}\@@href}%
\providecommand \@@href[1]{\endgroup#1\@@endlink}%
\providecommand \@sanitize@url [0]{\catcode `\\12\catcode `\$12\catcode
  `\&12\catcode `\#12\catcode `\^12\catcode `\_12\catcode `\%12\relax}%
\providecommand \@@startlink[1]{}%
\providecommand \@@endlink[0]{}%
\providecommand \url  [0]{\begingroup\@sanitize@url \@url }%
\providecommand \@url [1]{\endgroup\@href {#1}{\urlprefix }}%
\providecommand \urlprefix  [0]{URL }%
\providecommand \Eprint [0]{\href }%
\providecommand \doibase [0]{https://doi.org/}%
\providecommand \selectlanguage [0]{\@gobble}%
\providecommand \bibinfo  [0]{\@secondoftwo}%
\providecommand \bibfield  [0]{\@secondoftwo}%
\providecommand \translation [1]{[#1]}%
\providecommand \BibitemOpen [0]{}%
\providecommand \bibitemStop [0]{}%
\providecommand \bibitemNoStop [0]{.\EOS\space}%
\providecommand \EOS [0]{\spacefactor3000\relax}%
\providecommand \BibitemShut  [1]{\csname bibitem#1\endcsname}%
\let\auto@bib@innerbib\@empty
\bibitem [{\citenamefont {Dresselhaus}\ and\ \citenamefont
  {Dresselhaus}(1965)}]{Graphene1}%
  \BibitemOpen
  \bibfield  {author} {\bibinfo {author} {\bibfnamefont {G.}~\bibnamefont
  {Dresselhaus}}\ and\ \bibinfo {author} {\bibfnamefont {M.~S.}\ \bibnamefont
  {Dresselhaus}},\ }\bibfield  {title} {\bibinfo {title} {{Spin-Orbit
  Interaction in Graphite}},\ }\href {https://doi.org/10.1103/PhysRev.140.A401}
  {\bibfield  {journal} {\bibinfo  {journal} {Phys. Rev.}\ }\textbf {\bibinfo
  {volume} {140}},\ \bibinfo {pages} {A401} (\bibinfo {year}
  {1965})}\BibitemShut {NoStop}%
\bibitem [{\citenamefont {Kane}\ and\ \citenamefont {Mele}(2005)}]{Graphene2}%
  \BibitemOpen
  \bibfield  {author} {\bibinfo {author} {\bibfnamefont {C.~L.}\ \bibnamefont
  {Kane}}\ and\ \bibinfo {author} {\bibfnamefont {E.~J.}\ \bibnamefont
  {Mele}},\ }\bibfield  {title} {\bibinfo {title} {{Quantum Spin Hall Effect in
  Graphene}},\ }\href {https://doi.org/10.1103/PhysRevLett.95.226801}
  {\bibfield  {journal} {\bibinfo  {journal} {Phys. Rev. Lett.}\ }\textbf
  {\bibinfo {volume} {95}},\ \bibinfo {pages} {226801} (\bibinfo {year}
  {2005})}\BibitemShut {NoStop}%
\bibitem [{\citenamefont {Castro~Neto}\ \emph {et~al.}(2009)\citenamefont
  {Castro~Neto}, \citenamefont {Guinea}, \citenamefont {Peres}, \citenamefont
  {Novoselov},\ and\ \citenamefont {Geim}}]{CastroNeto2009Review}%
  \BibitemOpen
  \bibfield  {author} {\bibinfo {author} {\bibfnamefont {A.~H.}\ \bibnamefont
  {Castro~Neto}}, \bibinfo {author} {\bibfnamefont {F.}~\bibnamefont {Guinea}},
  \bibinfo {author} {\bibfnamefont {N.~M.~R.}\ \bibnamefont {Peres}}, \bibinfo
  {author} {\bibfnamefont {K.~S.}\ \bibnamefont {Novoselov}},\ and\ \bibinfo
  {author} {\bibfnamefont {A.~K.}\ \bibnamefont {Geim}},\ }\bibfield  {title}
  {\bibinfo {title} {The electronic properties of graphene},\ }\href
  {https://doi.org/10.1103/RevModPhys.81.109} {\bibfield  {journal} {\bibinfo
  {journal} {Rev. Mod. Phys.}\ }\textbf {\bibinfo {volume} {81}},\ \bibinfo
  {pages} {109} (\bibinfo {year} {2009})}\BibitemShut {NoStop}%
\bibitem [{\citenamefont {Bernevig}\ \emph {et~al.}(2006)\citenamefont
  {Bernevig}, \citenamefont {Hughes},\ and\ \citenamefont
  {Zhang}}]{bernevig2006BHZ}%
  \BibitemOpen
  \bibfield  {author} {\bibinfo {author} {\bibfnamefont {B.~A.}\ \bibnamefont
  {Bernevig}}, \bibinfo {author} {\bibfnamefont {T.~L.}\ \bibnamefont
  {Hughes}},\ and\ \bibinfo {author} {\bibfnamefont {S.-C.}\ \bibnamefont
  {Zhang}},\ }\bibfield  {title} {\bibinfo {title} {{Quantum spin Hall effect
  and topological phase transition in HgTe quantum wells}},\ }\href
  {https://doi.org/10.1126/science.1133734} {\bibfield  {journal} {\bibinfo
  {journal} {Science}\ }\textbf {\bibinfo {volume} {314}},\ \bibinfo {pages}
  {1757} (\bibinfo {year} {2006})}\BibitemShut {NoStop}%
\bibitem [{\citenamefont {Bernevig}\ and\ \citenamefont
  {Zhang}(2006)}]{Bernevig2006QSHE}%
  \BibitemOpen
  \bibfield  {author} {\bibinfo {author} {\bibfnamefont {B.~A.}\ \bibnamefont
  {Bernevig}}\ and\ \bibinfo {author} {\bibfnamefont {S.-C.}\ \bibnamefont
  {Zhang}},\ }\bibfield  {title} {\bibinfo {title} {Quantum spin hall effect},\
  }\href {https://doi.org/10.1103/PhysRevLett.96.106802} {\bibfield  {journal}
  {\bibinfo  {journal} {Phys. Rev. Lett.}\ }\textbf {\bibinfo {volume} {96}},\
  \bibinfo {pages} {106802} (\bibinfo {year} {2006})}\BibitemShut {NoStop}%
\bibitem [{\citenamefont {Hasan}\ and\ \citenamefont
  {Kane}(2010)}]{Kane2010Review}%
  \BibitemOpen
  \bibfield  {author} {\bibinfo {author} {\bibfnamefont {M.~Z.}\ \bibnamefont
  {Hasan}}\ and\ \bibinfo {author} {\bibfnamefont {C.~L.}\ \bibnamefont
  {Kane}},\ }\bibfield  {title} {\bibinfo {title} {Colloquium: Topological
  insulators},\ }\href {https://doi.org/10.1103/RevModPhys.82.3045} {\bibfield
  {journal} {\bibinfo  {journal} {Rev. Mod. Phys.}\ }\textbf {\bibinfo {volume}
  {82}},\ \bibinfo {pages} {3045} (\bibinfo {year} {2010})}\BibitemShut
  {NoStop}%
\bibitem [{\citenamefont {Qi}\ and\ \citenamefont
  {Zhang}(2011)}]{Zhang2011Review}%
  \BibitemOpen
  \bibfield  {author} {\bibinfo {author} {\bibfnamefont {X.-L.}\ \bibnamefont
  {Qi}}\ and\ \bibinfo {author} {\bibfnamefont {S.-C.}\ \bibnamefont {Zhang}},\
  }\bibfield  {title} {\bibinfo {title} {Topological insulators and
  superconductors},\ }\href {https://doi.org/10.1103/RevModPhys.83.1057}
  {\bibfield  {journal} {\bibinfo  {journal} {Rev. Mod. Phys.}\ }\textbf
  {\bibinfo {volume} {83}},\ \bibinfo {pages} {1057} (\bibinfo {year}
  {2011})}\BibitemShut {NoStop}%
\bibitem [{\citenamefont {Shen}(2012)}]{shen2012topological}%
  \BibitemOpen
  \bibfield  {author} {\bibinfo {author} {\bibfnamefont {S.-Q.}\ \bibnamefont
  {Shen}},\ }\href {https://doi.org/10.1007/978-3-642-32858-9} {\emph {\bibinfo
  {title} {{Topological insulators: Dirac Equation in Condensed Matter}}}},\
  \bibinfo {series} {Springer Series in Solid-State Sciences}, Vol.\ \bibinfo
  {volume} {174}\ (\bibinfo  {publisher} {Springer-Verlag, Berlin,
  Heidelberg},\ \bibinfo {year} {2012})\BibitemShut {NoStop}%
\bibitem [{\citenamefont {Bernevig}\ and\ \citenamefont
  {Hughes}(2013)}]{bernevig2013book}%
  \BibitemOpen
  \bibfield  {author} {\bibinfo {author} {\bibfnamefont {B.~A.}\ \bibnamefont
  {Bernevig}}\ and\ \bibinfo {author} {\bibfnamefont {T.~L.}\ \bibnamefont
  {Hughes}},\ }\href {https://press.princeton.edu/titles/10039.html} {\emph
  {\bibinfo {title} {Topological insulators and topological superconductors}}}\
  (\bibinfo  {publisher} {Princeton University Press},\ \bibinfo {year}
  {2013})\BibitemShut {NoStop}%
\bibitem [{\citenamefont {Fu}(2011)}]{Fu2011TCI}%
  \BibitemOpen
  \bibfield  {author} {\bibinfo {author} {\bibfnamefont {L.}~\bibnamefont
  {Fu}},\ }\bibfield  {title} {\bibinfo {title} {Topological crystalline
  insulators},\ }\href {https://doi.org/10.1103/PhysRevLett.106.106802}
  {\bibfield  {journal} {\bibinfo  {journal} {Phys. Rev. Lett.}\ }\textbf
  {\bibinfo {volume} {106}},\ \bibinfo {pages} {106802} (\bibinfo {year}
  {2011})}\BibitemShut {NoStop}%
\bibitem [{\citenamefont {Hsieh}\ \emph {et~al.}(2012)\citenamefont {Hsieh},
  \citenamefont {Lin}, \citenamefont {Liu}, \citenamefont {Duan}, \citenamefont
  {Bansil},\ and\ \citenamefont {Fu}}]{SnTe1}%
  \BibitemOpen
  \bibfield  {author} {\bibinfo {author} {\bibfnamefont {T.~H.}\ \bibnamefont
  {Hsieh}}, \bibinfo {author} {\bibfnamefont {H.}~\bibnamefont {Lin}}, \bibinfo
  {author} {\bibfnamefont {J.}~\bibnamefont {Liu}}, \bibinfo {author}
  {\bibfnamefont {W.}~\bibnamefont {Duan}}, \bibinfo {author} {\bibfnamefont
  {A.}~\bibnamefont {Bansil}},\ and\ \bibinfo {author} {\bibfnamefont
  {L.}~\bibnamefont {Fu}},\ }\bibfield  {title} {\bibinfo {title} {{Topological
  crystalline insulators in the SnTe material class}},\ }\href
  {https://doi.org/10.1038/ncomms1969} {\bibfield  {journal} {\bibinfo
  {journal} {Nat. Commun.}\ }\textbf {\bibinfo {volume} {3}},\ \bibinfo {pages}
  {982} (\bibinfo {year} {2012})}\BibitemShut {NoStop}%
\bibitem [{\citenamefont {Slager}\ \emph {et~al.}(2013)\citenamefont {Slager},
  \citenamefont {Mesaros}, \citenamefont {Juričić},\ and\ \citenamefont
  {Zaanen}}]{Slager2013SGClass}%
  \BibitemOpen
  \bibfield  {author} {\bibinfo {author} {\bibfnamefont {R.-J.}\ \bibnamefont
  {Slager}}, \bibinfo {author} {\bibfnamefont {A.}~\bibnamefont {Mesaros}},
  \bibinfo {author} {\bibfnamefont {V.}~\bibnamefont {Juričić}},\ and\
  \bibinfo {author} {\bibfnamefont {J.}~\bibnamefont {Zaanen}},\ }\bibfield
  {title} {\bibinfo {title} {{The space group classification of topological
  band-insulators}},\ }\href {https://doi.org/10.1038/nphys2513} {\bibfield
  {journal} {\bibinfo  {journal} {Nat. Phys.}\ }\textbf {\bibinfo {volume}
  {9}},\ \bibinfo {pages} {98} (\bibinfo {year} {2013})}\BibitemShut {NoStop}%
\bibitem [{\citenamefont {Wrasse}\ and\ \citenamefont {Schmidt}(2014)}]{PbSe1}%
  \BibitemOpen
  \bibfield  {author} {\bibinfo {author} {\bibfnamefont {E.~O.}\ \bibnamefont
  {Wrasse}}\ and\ \bibinfo {author} {\bibfnamefont {T.~M.}\ \bibnamefont
  {Schmidt}},\ }\bibfield  {title} {\bibinfo {title} {{Prediction of
  Two-Dimensional Topological Crystalline Insulator in PbSe Monolayer}},\
  }\href {https://doi.org/10.1021/nl502481f} {\bibfield  {journal} {\bibinfo
  {journal} {Nano Letters}\ }\textbf {\bibinfo {volume} {14}},\ \bibinfo
  {pages} {5717–5720} (\bibinfo {year} {2014})}\BibitemShut {NoStop}%
\bibitem [{\citenamefont {Ando}\ and\ \citenamefont
  {Fu}(2015)}]{Fu2015ReviewTCI}%
  \BibitemOpen
  \bibfield  {author} {\bibinfo {author} {\bibfnamefont {Y.}~\bibnamefont
  {Ando}}\ and\ \bibinfo {author} {\bibfnamefont {L.}~\bibnamefont {Fu}},\
  }\bibfield  {title} {\bibinfo {title} {{Topological Crystalline Insulators
  and Topological Superconductors: From Concepts to Materials}},\ }\href
  {https://doi.org/10.1146/annurev-conmatphys-031214-014501} {\bibfield
  {journal} {\bibinfo  {journal} {Annu. Rev. Condens. Matter Phys.}\ }\textbf
  {\bibinfo {volume} {6}},\ \bibinfo {pages} {361} (\bibinfo {year}
  {2015})}\BibitemShut {NoStop}%
\bibitem [{\citenamefont {Kruthoff}\ \emph {et~al.}(2017)\citenamefont
  {Kruthoff}, \citenamefont {de~Boer}, \citenamefont {van Wezel}, \citenamefont
  {Kane},\ and\ \citenamefont {Slager}}]{Slager2017TopoClassCombina}%
  \BibitemOpen
  \bibfield  {author} {\bibinfo {author} {\bibfnamefont {J.}~\bibnamefont
  {Kruthoff}}, \bibinfo {author} {\bibfnamefont {J.}~\bibnamefont {de~Boer}},
  \bibinfo {author} {\bibfnamefont {J.}~\bibnamefont {van Wezel}}, \bibinfo
  {author} {\bibfnamefont {C.~L.}\ \bibnamefont {Kane}},\ and\ \bibinfo
  {author} {\bibfnamefont {R.-J.}\ \bibnamefont {Slager}},\ }\bibfield  {title}
  {\bibinfo {title} {Topological classification of crystalline insulators
  through band structure combinatorics},\ }\href
  {https://doi.org/10.1103/PhysRevX.7.041069} {\bibfield  {journal} {\bibinfo
  {journal} {Phys. Rev. X}\ }\textbf {\bibinfo {volume} {7}},\ \bibinfo {pages}
  {041069} (\bibinfo {year} {2017})}\BibitemShut {NoStop}%
\bibitem [{\citenamefont {Benalcazar}\ \emph
  {et~al.}(2017{\natexlab{a}})\citenamefont {Benalcazar}, \citenamefont
  {Bernevig},\ and\ \citenamefont {Hughes}}]{Benalcazar2017Multipole}%
  \BibitemOpen
  \bibfield  {author} {\bibinfo {author} {\bibfnamefont {W.~A.}\ \bibnamefont
  {Benalcazar}}, \bibinfo {author} {\bibfnamefont {B.~A.}\ \bibnamefont
  {Bernevig}},\ and\ \bibinfo {author} {\bibfnamefont {T.~L.}\ \bibnamefont
  {Hughes}},\ }\bibfield  {title} {\bibinfo {title} {Electric multipole
  moments, topological multipole moment pumping, and chiral hinge states in
  crystalline insulators},\ }\href {https://doi.org/10.1103/PhysRevB.96.245115}
  {\bibfield  {journal} {\bibinfo  {journal} {Phys. Rev. B}\ }\textbf {\bibinfo
  {volume} {96}},\ \bibinfo {pages} {245115} (\bibinfo {year}
  {2017}{\natexlab{a}})}\BibitemShut {NoStop}%
\bibitem [{\citenamefont {Benalcazar}\ \emph
  {et~al.}(2017{\natexlab{b}})\citenamefont {Benalcazar}, \citenamefont
  {Bernevig},\ and\ \citenamefont {Hughes}}]{Benalcazar2017QEML}%
  \BibitemOpen
  \bibfield  {author} {\bibinfo {author} {\bibfnamefont {W.~A.}\ \bibnamefont
  {Benalcazar}}, \bibinfo {author} {\bibfnamefont {B.~A.}\ \bibnamefont
  {Bernevig}},\ and\ \bibinfo {author} {\bibfnamefont {T.~L.}\ \bibnamefont
  {Hughes}},\ }\bibfield  {title} {\bibinfo {title} {Quantized electric
  multipole insulators},\ }\href {https://doi.org/10.1126/science.aah6442}
  {\bibfield  {journal} {\bibinfo  {journal} {Science}\ }\textbf {\bibinfo
  {volume} {357}},\ \bibinfo {pages} {61} (\bibinfo {year}
  {2017}{\natexlab{b}})}\BibitemShut {NoStop}%
\bibitem [{\citenamefont {Langbehn}\ \emph {et~al.}(2017)\citenamefont
  {Langbehn}, \citenamefont {Peng}, \citenamefont {Trifunovic}, \citenamefont
  {von Oppen},\ and\ \citenamefont {Brouwer}}]{Langbehn2017HOTI}%
  \BibitemOpen
  \bibfield  {author} {\bibinfo {author} {\bibfnamefont {J.}~\bibnamefont
  {Langbehn}}, \bibinfo {author} {\bibfnamefont {Y.}~\bibnamefont {Peng}},
  \bibinfo {author} {\bibfnamefont {L.}~\bibnamefont {Trifunovic}}, \bibinfo
  {author} {\bibfnamefont {F.}~\bibnamefont {von Oppen}},\ and\ \bibinfo
  {author} {\bibfnamefont {P.~W.}\ \bibnamefont {Brouwer}},\ }\bibfield
  {title} {\bibinfo {title} {Reflection-symmetric second-order topological
  insulators and superconductors},\ }\href
  {https://doi.org/10.1103/PhysRevLett.119.246401} {\bibfield  {journal}
  {\bibinfo  {journal} {Phys. Rev. Lett.}\ }\textbf {\bibinfo {volume} {119}},\
  \bibinfo {pages} {246401} (\bibinfo {year} {2017})}\BibitemShut {NoStop}%
\bibitem [{\citenamefont {Schindler}\ \emph {et~al.}(2018)\citenamefont
  {Schindler}, \citenamefont {Cook}, \citenamefont {Vergniory}, \citenamefont
  {Wang}, \citenamefont {Parkin}, \citenamefont {Bernevig},\ and\ \citenamefont
  {Neupert}}]{Schindler2018HOTI}%
  \BibitemOpen
  \bibfield  {author} {\bibinfo {author} {\bibfnamefont {F.}~\bibnamefont
  {Schindler}}, \bibinfo {author} {\bibfnamefont {A.~M.}\ \bibnamefont {Cook}},
  \bibinfo {author} {\bibfnamefont {M.~G.}\ \bibnamefont {Vergniory}}, \bibinfo
  {author} {\bibfnamefont {Z.}~\bibnamefont {Wang}}, \bibinfo {author}
  {\bibfnamefont {S.~S.~P.}\ \bibnamefont {Parkin}}, \bibinfo {author}
  {\bibfnamefont {B.~A.}\ \bibnamefont {Bernevig}},\ and\ \bibinfo {author}
  {\bibfnamefont {T.}~\bibnamefont {Neupert}},\ }\bibfield  {title} {\bibinfo
  {title} {Higher-order topological insulators},\ }\href
  {https://doi.org/10.1126/sciadv.aat0346} {\bibfield  {journal} {\bibinfo
  {journal} {Sci. Adv.}\ }\textbf {\bibinfo {volume} {4}},\ \bibinfo {pages}
  {eaat0346} (\bibinfo {year} {2018})}\BibitemShut {NoStop}%
\bibitem [{\citenamefont {Wan}\ \emph {et~al.}(2011)\citenamefont {Wan},
  \citenamefont {Turner}, \citenamefont {Vishwanath},\ and\ \citenamefont
  {Savrasov}}]{Wan2011WSM}%
  \BibitemOpen
  \bibfield  {author} {\bibinfo {author} {\bibfnamefont {X.}~\bibnamefont
  {Wan}}, \bibinfo {author} {\bibfnamefont {A.~M.}\ \bibnamefont {Turner}},
  \bibinfo {author} {\bibfnamefont {A.}~\bibnamefont {Vishwanath}},\ and\
  \bibinfo {author} {\bibfnamefont {S.~Y.}\ \bibnamefont {Savrasov}},\
  }\bibfield  {title} {\bibinfo {title} {Topological semimetal and fermi-arc
  surface states in the electronic structure of pyrochlore iridates},\ }\href
  {https://doi.org/10.1103/PhysRevB.83.205101} {\bibfield  {journal} {\bibinfo
  {journal} {Phys. Rev. B}\ }\textbf {\bibinfo {volume} {83}},\ \bibinfo
  {pages} {205101} (\bibinfo {year} {2011})}\BibitemShut {NoStop}%
\bibitem [{\citenamefont {Xu}\ \emph {et~al.}(2015)\citenamefont {Xu},
  \citenamefont {Belopolski}, \citenamefont {Alidoust}, \citenamefont
  {Neupane}, \citenamefont {Bian}, \citenamefont {Zhang}, \citenamefont
  {Sankar}, \citenamefont {Chang}, \citenamefont {Yuan}, \citenamefont {Lee},
  \citenamefont {Huang}, \citenamefont {Zheng}, \citenamefont {Ma},
  \citenamefont {Sanchez}, \citenamefont {Wang}, \citenamefont {Bansil},
  \citenamefont {Chou}, \citenamefont {Shibayev}, \citenamefont {Lin},
  \citenamefont {Jia},\ and\ \citenamefont {Hasan}}]{Xu2015WSMExp}%
  \BibitemOpen
  \bibfield  {author} {\bibinfo {author} {\bibfnamefont {S.-Y.}\ \bibnamefont
  {Xu}}, \bibinfo {author} {\bibfnamefont {I.}~\bibnamefont {Belopolski}},
  \bibinfo {author} {\bibfnamefont {N.}~\bibnamefont {Alidoust}}, \bibinfo
  {author} {\bibfnamefont {M.}~\bibnamefont {Neupane}}, \bibinfo {author}
  {\bibfnamefont {G.}~\bibnamefont {Bian}}, \bibinfo {author} {\bibfnamefont
  {C.}~\bibnamefont {Zhang}}, \bibinfo {author} {\bibfnamefont
  {R.}~\bibnamefont {Sankar}}, \bibinfo {author} {\bibfnamefont
  {G.}~\bibnamefont {Chang}}, \bibinfo {author} {\bibfnamefont
  {Z.}~\bibnamefont {Yuan}}, \bibinfo {author} {\bibfnamefont {C.-C.}\
  \bibnamefont {Lee}}, \bibinfo {author} {\bibfnamefont {S.-M.}\ \bibnamefont
  {Huang}}, \bibinfo {author} {\bibfnamefont {H.}~\bibnamefont {Zheng}},
  \bibinfo {author} {\bibfnamefont {J.}~\bibnamefont {Ma}}, \bibinfo {author}
  {\bibfnamefont {D.~S.}\ \bibnamefont {Sanchez}}, \bibinfo {author}
  {\bibfnamefont {B.}~\bibnamefont {Wang}}, \bibinfo {author} {\bibfnamefont
  {A.}~\bibnamefont {Bansil}}, \bibinfo {author} {\bibfnamefont
  {F.}~\bibnamefont {Chou}}, \bibinfo {author} {\bibfnamefont {P.~P.}\
  \bibnamefont {Shibayev}}, \bibinfo {author} {\bibfnamefont {H.}~\bibnamefont
  {Lin}}, \bibinfo {author} {\bibfnamefont {S.}~\bibnamefont {Jia}},\ and\
  \bibinfo {author} {\bibfnamefont {M.~Z.}\ \bibnamefont {Hasan}},\ }\bibfield
  {title} {\bibinfo {title} {{Discovery of a Weyl fermion semimetal and
  topological Fermi arcs}},\ }\href {https://doi.org/10.1126/science.aaa9297}
  {\bibfield  {journal} {\bibinfo  {journal} {Science}\ }\textbf {\bibinfo
  {volume} {349}},\ \bibinfo {pages} {613} (\bibinfo {year}
  {2015})}\BibitemShut {NoStop}%
\bibitem [{\citenamefont {Yue}\ \emph {et~al.}(2017)\citenamefont {Yue},
  \citenamefont {Xue}, \citenamefont {Liu}, \citenamefont {Wang},\ and\
  \citenamefont {Gu}}]{device1}%
  \BibitemOpen
  \bibfield  {author} {\bibinfo {author} {\bibfnamefont {Z.}~\bibnamefont
  {Yue}}, \bibinfo {author} {\bibfnamefont {G.}~\bibnamefont {Xue}}, \bibinfo
  {author} {\bibfnamefont {J.}~\bibnamefont {Liu}}, \bibinfo {author}
  {\bibfnamefont {Y.}~\bibnamefont {Wang}},\ and\ \bibinfo {author}
  {\bibfnamefont {M.}~\bibnamefont {Gu}},\ }\bibfield  {title} {\bibinfo
  {title} {Nanometric holograms based on a topological insulator material},\
  }\href {https://doi.org/10.1038/ncomms15354} {\bibfield  {journal} {\bibinfo
  {journal} {Nat. Commun.}\ }\textbf {\bibinfo {volume} {8}},\ \bibinfo {pages}
  {15354} (\bibinfo {year} {2017})}\BibitemShut {NoStop}%
\bibitem [{\citenamefont {Goi}\ \emph {et~al.}(2018)\citenamefont {Goi},
  \citenamefont {Yue}, \citenamefont {Cumming},\ and\ \citenamefont
  {Gu}}]{device2}%
  \BibitemOpen
  \bibfield  {author} {\bibinfo {author} {\bibfnamefont {E.}~\bibnamefont
  {Goi}}, \bibinfo {author} {\bibfnamefont {Z.}~\bibnamefont {Yue}}, \bibinfo
  {author} {\bibfnamefont {B.~P.}\ \bibnamefont {Cumming}},\ and\ \bibinfo
  {author} {\bibfnamefont {M.}~\bibnamefont {Gu}},\ }\bibfield  {title}
  {\bibinfo {title} {{Observation of Type I Photonic Weyl Points in Optical
  Frequencies}},\ }\href {https://doi.org/10.1002/lpor.201700271} {\bibfield
  {journal} {\bibinfo  {journal} {Laser Photon. Rev.}\ }\textbf {\bibinfo
  {volume} {12}},\ \bibinfo {pages} {1700271} (\bibinfo {year}
  {2018})}\BibitemShut {NoStop}%
\bibitem [{\citenamefont {Yue}\ \emph {et~al.}(2018)\citenamefont {Yue},
  \citenamefont {Jiang}, \citenamefont {Zhu}, \citenamefont {Chen},
  \citenamefont {Sun},\ and\ \citenamefont {Zhang}}]{device4}%
  \BibitemOpen
  \bibfield  {author} {\bibinfo {author} {\bibfnamefont {C.}~\bibnamefont
  {Yue}}, \bibinfo {author} {\bibfnamefont {S.}~\bibnamefont {Jiang}}, \bibinfo
  {author} {\bibfnamefont {H.}~\bibnamefont {Zhu}}, \bibinfo {author}
  {\bibfnamefont {L.}~\bibnamefont {Chen}}, \bibinfo {author} {\bibfnamefont
  {Q.}~\bibnamefont {Sun}},\ and\ \bibinfo {author} {\bibfnamefont {D.~W.}\
  \bibnamefont {Zhang}},\ }\bibfield  {title} {\bibinfo {title} {Device
  applications of synthetic topological insulator nanostructures},\ }\href
  {https://doi.org/10.3390/electronics7100225} {\bibfield  {journal} {\bibinfo
  {journal} {Electronics}\ }\textbf {\bibinfo {volume} {7}},\ \bibinfo {pages}
  {225} (\bibinfo {year} {2018})}\BibitemShut {NoStop}%
\bibitem [{\citenamefont {Fan}\ and\ \citenamefont {Wang}(2016)}]{device3}%
  \BibitemOpen
  \bibfield  {author} {\bibinfo {author} {\bibfnamefont {Y.}~\bibnamefont
  {Fan}}\ and\ \bibinfo {author} {\bibfnamefont {K.~L.}\ \bibnamefont {Wang}},\
  }\bibfield  {title} {\bibinfo {title} {Spintronics based on topological
  insulators},\ }\href {https://doi.org/10.1142/S2010324716400014} {\bibfield
  {journal} {\bibinfo  {journal} {SPIN}\ }\textbf {\bibinfo {volume} {6}},\
  \bibinfo {pages} {1640001} (\bibinfo {year} {2016})}\BibitemShut {NoStop}%
\bibitem [{\citenamefont {Wilson}(1974)}]{wilson1974confinement}%
  \BibitemOpen
  \bibfield  {author} {\bibinfo {author} {\bibfnamefont {K.~G.}\ \bibnamefont
  {Wilson}},\ }\bibfield  {title} {\bibinfo {title} {Confinement of quarks},\
  }\href {https://doi.org/10.1103/PhysRevD.10.2445} {\bibfield  {journal}
  {\bibinfo  {journal} {Phys. Rev. D}\ }\textbf {\bibinfo {volume} {10}},\
  \bibinfo {pages} {2445} (\bibinfo {year} {1974})}\BibitemShut {NoStop}%
\bibitem [{\citenamefont {Kogut}\ and\ \citenamefont
  {Susskind}(1975)}]{Kogut1975}%
  \BibitemOpen
  \bibfield  {author} {\bibinfo {author} {\bibfnamefont {J.~B.}\ \bibnamefont
  {Kogut}}\ and\ \bibinfo {author} {\bibfnamefont {L.}~\bibnamefont
  {Susskind}},\ }\bibfield  {title} {\bibinfo {title} {{Hamiltonian formulation
  of Wilson's lattice gauge theories}},\ }\href
  {https://doi.org/10.1103/PhysRevD.11.395} {\bibfield  {journal} {\bibinfo
  {journal} {Phys. Rev. D}\ }\textbf {\bibinfo {volume} {11}},\ \bibinfo
  {pages} {395} (\bibinfo {year} {1975})}\BibitemShut {NoStop}%
\bibitem [{\citenamefont {Nielsen}\ and\ \citenamefont
  {Ninomiya}(1981{\natexlab{a}})}]{NIELSEN1981219}%
  \BibitemOpen
  \bibfield  {author} {\bibinfo {author} {\bibfnamefont {H.~B.}\ \bibnamefont
  {Nielsen}}\ and\ \bibinfo {author} {\bibfnamefont {M.}~\bibnamefont
  {Ninomiya}},\ }\bibfield  {title} {\bibinfo {title} {A no-go theorem for
  regularizing chiral fermions},\ }\href
  {https://doi.org/10.1016/0370-2693(81)91026-1} {\bibfield  {journal}
  {\bibinfo  {journal} {Phys. Lett. B}\ }\textbf {\bibinfo {volume} {105}},\
  \bibinfo {pages} {219} (\bibinfo {year} {1981}{\natexlab{a}})}\BibitemShut
  {NoStop}%
\bibitem [{\citenamefont {Nielsen}\ and\ \citenamefont
  {Ninomiya}(1981{\natexlab{b}})}]{nielsen1981absence}%
  \BibitemOpen
  \bibfield  {author} {\bibinfo {author} {\bibfnamefont {H.~B.}\ \bibnamefont
  {Nielsen}}\ and\ \bibinfo {author} {\bibfnamefont {M.}~\bibnamefont
  {Ninomiya}},\ }\bibfield  {title} {\bibinfo {title} {{Absence of neutrinos on
  a lattice: (I). Proof by homotopy theory}},\ }\href
  {https://doi.org/10.1016/0550-3213(81)90361-8} {\bibfield  {journal}
  {\bibinfo  {journal} {Nucl. Phys. B}\ }\textbf {\bibinfo {volume} {185}},\
  \bibinfo {pages} {20} (\bibinfo {year} {1981}{\natexlab{b}})}\BibitemShut
  {NoStop}%
\bibitem [{\citenamefont {Candido}\ \emph {et~al.}(2018)\citenamefont
  {Candido}, \citenamefont {Kharitonov}, \citenamefont {Egues},\ and\
  \citenamefont {Hankiewicz}}]{Denis2018Paradoxical}%
  \BibitemOpen
  \bibfield  {author} {\bibinfo {author} {\bibfnamefont {D.~R.}\ \bibnamefont
  {Candido}}, \bibinfo {author} {\bibfnamefont {M.}~\bibnamefont {Kharitonov}},
  \bibinfo {author} {\bibfnamefont {J.~C.}\ \bibnamefont {Egues}},\ and\
  \bibinfo {author} {\bibfnamefont {E.~M.}\ \bibnamefont {Hankiewicz}},\
  }\bibfield  {title} {\bibinfo {title} {{Paradoxical extension of the edge
  states across the topological phase transition due to emergent approximate
  chiral symmetry in a quantum anomalous Hall system}},\ }\href
  {https://doi.org/10.1103/PhysRevB.98.161111} {\bibfield  {journal} {\bibinfo
  {journal} {Phys. Rev. B}\ }\textbf {\bibinfo {volume} {98}},\ \bibinfo
  {pages} {161111(R)} (\bibinfo {year} {2018})}\BibitemShut {NoStop}%
\bibitem [{\citenamefont {Tworzyd\l{}o}\ \emph {et~al.}(2008)\citenamefont
  {Tworzyd\l{}o}, \citenamefont {Groth},\ and\ \citenamefont
  {Beenakker}}]{Beenakker2008FiniteDiff}%
  \BibitemOpen
  \bibfield  {author} {\bibinfo {author} {\bibfnamefont {J.}~\bibnamefont
  {Tworzyd\l{}o}}, \bibinfo {author} {\bibfnamefont {C.~W.}\ \bibnamefont
  {Groth}},\ and\ \bibinfo {author} {\bibfnamefont {C.~W.~J.}\ \bibnamefont
  {Beenakker}},\ }\bibfield  {title} {\bibinfo {title} {Finite difference
  method for transport properties of massless dirac fermions},\ }\href
  {https://doi.org/10.1103/PhysRevB.78.235438} {\bibfield  {journal} {\bibinfo
  {journal} {Phys. Rev. B}\ }\textbf {\bibinfo {volume} {78}},\ \bibinfo
  {pages} {235438} (\bibinfo {year} {2008})}\BibitemShut {NoStop}%
\bibitem [{\citenamefont {Hern\'andez}\ and\ \citenamefont
  {Lewenkopf}(2012)}]{AlexisCaio2012}%
  \BibitemOpen
  \bibfield  {author} {\bibinfo {author} {\bibfnamefont {A.~R.}\ \bibnamefont
  {Hern\'andez}}\ and\ \bibinfo {author} {\bibfnamefont {C.~H.}\ \bibnamefont
  {Lewenkopf}},\ }\bibfield  {title} {\bibinfo {title} {{Finite-difference
  method for transport of two-dimensional massless Dirac fermions in a ribbon
  geometry}},\ }\href {https://doi.org/10.1103/PhysRevB.86.155439} {\bibfield
  {journal} {\bibinfo  {journal} {Phys. Rev. B}\ }\textbf {\bibinfo {volume}
  {86}},\ \bibinfo {pages} {155439} (\bibinfo {year} {2012})}\BibitemShut
  {NoStop}%
\bibitem [{\citenamefont {Zhou}\ \emph {et~al.}(2017)\citenamefont {Zhou},
  \citenamefont {Jiang}, \citenamefont {Xie},\ and\ \citenamefont
  {Sun}}]{zhou2016LatticeModel}%
  \BibitemOpen
  \bibfield  {author} {\bibinfo {author} {\bibfnamefont {Y.-F.}\ \bibnamefont
  {Zhou}}, \bibinfo {author} {\bibfnamefont {H.}~\bibnamefont {Jiang}},
  \bibinfo {author} {\bibfnamefont {X.~C.}\ \bibnamefont {Xie}},\ and\ \bibinfo
  {author} {\bibfnamefont {Q.-F.}\ \bibnamefont {Sun}},\ }\bibfield  {title}
  {\bibinfo {title} {Two-dimensional lattice model for the surface states of
  topological insulators},\ }\href {https://doi.org/10.1103/PhysRevB.95.245137}
  {\bibfield  {journal} {\bibinfo  {journal} {Phys. Rev. B}\ }\textbf {\bibinfo
  {volume} {95}},\ \bibinfo {pages} {245137} (\bibinfo {year}
  {2017})}\BibitemShut {NoStop}%
\bibitem [{\citenamefont {Gruji\ifmmode~\acute{c}\else \'{c}\fi{}}\ \emph
  {et~al.}(2011)\citenamefont {Gruji\ifmmode~\acute{c}\else \'{c}\fi{}},
  \citenamefont {Zarenia}, \citenamefont {Chaves}, \citenamefont
  {Tadi\ifmmode~\acute{c}\else \'{c}\fi{}}, \citenamefont {Farias},\ and\
  \citenamefont {Peeters}}]{Peeters2011CircularGrapheneDot}%
  \BibitemOpen
  \bibfield  {author} {\bibinfo {author} {\bibfnamefont {M.}~\bibnamefont
  {Gruji\ifmmode~\acute{c}\else \'{c}\fi{}}}, \bibinfo {author} {\bibfnamefont
  {M.}~\bibnamefont {Zarenia}}, \bibinfo {author} {\bibfnamefont
  {A.}~\bibnamefont {Chaves}}, \bibinfo {author} {\bibfnamefont
  {M.}~\bibnamefont {Tadi\ifmmode~\acute{c}\else \'{c}\fi{}}}, \bibinfo
  {author} {\bibfnamefont {G.~A.}\ \bibnamefont {Farias}},\ and\ \bibinfo
  {author} {\bibfnamefont {F.~M.}\ \bibnamefont {Peeters}},\ }\bibfield
  {title} {\bibinfo {title} {{Electronic and optical properties of a circular
  graphene quantum dot in a magnetic field: Influence of the boundary
  conditions}},\ }\href {https://doi.org/10.1103/PhysRevB.84.205441} {\bibfield
   {journal} {\bibinfo  {journal} {Phys. Rev. B}\ }\textbf {\bibinfo {volume}
  {84}},\ \bibinfo {pages} {205441} (\bibinfo {year} {2011})}\BibitemShut
  {NoStop}%
\bibitem [{\citenamefont {Zarenia}\ \emph {et~al.}(2011)\citenamefont
  {Zarenia}, \citenamefont {Chaves}, \citenamefont {Farias},\ and\
  \citenamefont {Peeters}}]{Peeters2011GrapheneDots}%
  \BibitemOpen
  \bibfield  {author} {\bibinfo {author} {\bibfnamefont {M.}~\bibnamefont
  {Zarenia}}, \bibinfo {author} {\bibfnamefont {A.}~\bibnamefont {Chaves}},
  \bibinfo {author} {\bibfnamefont {G.~A.}\ \bibnamefont {Farias}},\ and\
  \bibinfo {author} {\bibfnamefont {F.~M.}\ \bibnamefont {Peeters}},\
  }\bibfield  {title} {\bibinfo {title} {{Energy levels of triangular and
  hexagonal graphene quantum dots: A comparative study between the
  tight-binding and Dirac equation approach}},\ }\href
  {https://doi.org/10.1103/PhysRevB.84.245403} {\bibfield  {journal} {\bibinfo
  {journal} {Phys. Rev. B}\ }\textbf {\bibinfo {volume} {84}},\ \bibinfo
  {pages} {245403} (\bibinfo {year} {2011})}\BibitemShut {NoStop}%
\bibitem [{\citenamefont {Mirzakhani}\ \emph {et~al.}(2016)\citenamefont
  {Mirzakhani}, \citenamefont {Zarenia}, \citenamefont {da~Costa},
  \citenamefont {Ketabi},\ and\ \citenamefont
  {Peeters}}]{Peeters2016ABCgraphene}%
  \BibitemOpen
  \bibfield  {author} {\bibinfo {author} {\bibfnamefont {M.}~\bibnamefont
  {Mirzakhani}}, \bibinfo {author} {\bibfnamefont {M.}~\bibnamefont {Zarenia}},
  \bibinfo {author} {\bibfnamefont {D.~R.}\ \bibnamefont {da~Costa}}, \bibinfo
  {author} {\bibfnamefont {S.~A.}\ \bibnamefont {Ketabi}},\ and\ \bibinfo
  {author} {\bibfnamefont {F.~M.}\ \bibnamefont {Peeters}},\ }\bibfield
  {title} {\bibinfo {title} {{Energy levels of ABC-stacked trilayer graphene
  quantum dots with infinite-mass boundary conditions}},\ }\href
  {https://doi.org/10.1103/PhysRevB.94.165423} {\bibfield  {journal} {\bibinfo
  {journal} {Phys. Rev. B}\ }\textbf {\bibinfo {volume} {94}},\ \bibinfo
  {pages} {165423} (\bibinfo {year} {2016})}\BibitemShut {NoStop}%
\bibitem [{\citenamefont {McCann}\ and\ \citenamefont
  {Fal’ko}(2004)}]{McCannFalko2004}%
  \BibitemOpen
  \bibfield  {author} {\bibinfo {author} {\bibfnamefont {E.}~\bibnamefont
  {McCann}}\ and\ \bibinfo {author} {\bibfnamefont {V.~I.}\ \bibnamefont
  {Fal’ko}},\ }\bibfield  {title} {\bibinfo {title} {{Symmetry of boundary
  conditions of the Dirac equation for electrons in carbon nanotubes}},\ }\href
  {https://doi.org/10.1088/0953-8984/16/13/016} {\bibfield  {journal} {\bibinfo
   {journal} {J. Phys. Condens. Matter}\ }\textbf {\bibinfo {volume} {16}},\
  \bibinfo {pages} {2371} (\bibinfo {year} {2004})}\BibitemShut {NoStop}%
\bibitem [{\citenamefont {Akhmerov}\ and\ \citenamefont
  {Beenakker}(2008)}]{Akhmerov2008Boundary}%
  \BibitemOpen
  \bibfield  {author} {\bibinfo {author} {\bibfnamefont {A.~R.}\ \bibnamefont
  {Akhmerov}}\ and\ \bibinfo {author} {\bibfnamefont {C.~W.~J.}\ \bibnamefont
  {Beenakker}},\ }\bibfield  {title} {\bibinfo {title} {{Boundary conditions
  for Dirac fermions on a terminated honeycomb lattice}},\ }\href
  {https://doi.org/10.1103/PhysRevB.77.085423} {\bibfield  {journal} {\bibinfo
  {journal} {Phys. Rev. B}\ }\textbf {\bibinfo {volume} {77}},\ \bibinfo
  {pages} {085423} (\bibinfo {year} {2008})}\BibitemShut {NoStop}%
\bibitem [{\citenamefont {Berry}\ and\ \citenamefont
  {Mondragon}(1987)}]{BerryMondragon1987}%
  \BibitemOpen
  \bibfield  {author} {\bibinfo {author} {\bibfnamefont {M.~V.}\ \bibnamefont
  {Berry}}\ and\ \bibinfo {author} {\bibfnamefont {R.~J.}\ \bibnamefont
  {Mondragon}},\ }\bibfield  {title} {\bibinfo {title} {Neutrino billiards:
  time-reversal symmetry-breaking without magnetic fields},\ }\href
  {https://doi.org/10.1098/rspa.1987.0080} {\bibfield  {journal} {\bibinfo
  {journal} {Proc. Royal Soc. Lond.}\ }\textbf {\bibinfo {volume} {412}},\
  \bibinfo {pages} {53} (\bibinfo {year} {1987})}\BibitemShut {NoStop}%
\bibitem [{\citenamefont {Brey}\ and\ \citenamefont
  {Fertig}(2006)}]{BreyFertig2006}%
  \BibitemOpen
  \bibfield  {author} {\bibinfo {author} {\bibfnamefont {L.}~\bibnamefont
  {Brey}}\ and\ \bibinfo {author} {\bibfnamefont {H.~A.}\ \bibnamefont
  {Fertig}},\ }\bibfield  {title} {\bibinfo {title} {{Electronic states of
  graphene nanoribbons studied with the Dirac equation}},\ }\href
  {https://doi.org/10.1103/PhysRevB.73.235411} {\bibfield  {journal} {\bibinfo
  {journal} {Phys. Rev. B}\ }\textbf {\bibinfo {volume} {73}},\ \bibinfo
  {pages} {235411} (\bibinfo {year} {2006})}\BibitemShut {NoStop}%
\bibitem [{\citenamefont {Messias~de Resende}\ \emph
  {et~al.}(2017)\citenamefont {Messias~de Resende}, \citenamefont {de~Lima},
  \citenamefont {Miwa}, \citenamefont {Vernek},\ and\ \citenamefont
  {Ferreira}}]{Bruno2017FDP}%
  \BibitemOpen
  \bibfield  {author} {\bibinfo {author} {\bibfnamefont {B.}~\bibnamefont
  {Messias~de Resende}}, \bibinfo {author} {\bibfnamefont {F.~C.}\ \bibnamefont
  {de~Lima}}, \bibinfo {author} {\bibfnamefont {R.~H.}\ \bibnamefont {Miwa}},
  \bibinfo {author} {\bibfnamefont {E.}~\bibnamefont {Vernek}},\ and\ \bibinfo
  {author} {\bibfnamefont {G.~J.}\ \bibnamefont {Ferreira}},\ }\bibfield
  {title} {\bibinfo {title} {{Confinement and fermion doubling problem in
  Dirac-like Hamiltonians}},\ }\href
  {https://doi.org/10.1103/PhysRevB.96.161113} {\bibfield  {journal} {\bibinfo
  {journal} {Phys. Rev. B}\ }\textbf {\bibinfo {volume} {96}},\ \bibinfo
  {pages} {161113(R)} (\bibinfo {year} {2017})}\BibitemShut {NoStop}%
\bibitem [{\citenamefont {Varjas}\ \emph {et~al.}(2018)\citenamefont {Varjas},
  \citenamefont {Rosdahl},\ and\ \citenamefont {Akhmerov}}]{Qsymm2018}%
  \BibitemOpen
  \bibfield  {author} {\bibinfo {author} {\bibfnamefont {D.}~\bibnamefont
  {Varjas}}, \bibinfo {author} {\bibfnamefont {T.~O.}\ \bibnamefont
  {Rosdahl}},\ and\ \bibinfo {author} {\bibfnamefont {A.~R.}\ \bibnamefont
  {Akhmerov}},\ }\bibfield  {title} {\bibinfo {title} {{Qsymm: algorithmic
  symmetry finding and symmetric Hamiltonian generation}},\ }\href
  {https://doi.org/10.1088/1367-2630/aadf67} {\bibfield  {journal} {\bibinfo
  {journal} {New J. of Phys.}\ }\textbf {\bibinfo {volume} {20}},\ \bibinfo
  {pages} {093026} (\bibinfo {year} {2018})}\BibitemShut {NoStop}%
\bibitem [{\citenamefont {Groth}\ \emph {et~al.}(2014)\citenamefont {Groth},
  \citenamefont {Wimmer}, \citenamefont {Akhmerov},\ and\ \citenamefont
  {Waintal}}]{kwant}%
  \BibitemOpen
  \bibfield  {author} {\bibinfo {author} {\bibfnamefont {C.~W.}\ \bibnamefont
  {Groth}}, \bibinfo {author} {\bibfnamefont {M.}~\bibnamefont {Wimmer}},
  \bibinfo {author} {\bibfnamefont {A.~R.}\ \bibnamefont {Akhmerov}},\ and\
  \bibinfo {author} {\bibfnamefont {X.}~\bibnamefont {Waintal}},\ }\bibfield
  {title} {\bibinfo {title} {Kwant: a software package for quantum transport},\
  }\href {https://doi.org/doi:10.1088/1367-2630/16/6/063065} {\bibfield
  {journal} {\bibinfo  {journal} {New J. Phys.}\ }\textbf {\bibinfo {volume}
  {16}},\ \bibinfo {pages} {063065} (\bibinfo {year} {2014})}\BibitemShut
  {NoStop}%
\bibitem [{SM()}]{SM}%
  \BibitemOpen
  \href@noop {} {}\bibinfo {note} {{Supplemental material is available at the
  arXiv repository, which includes (i) the DFT/VASP raw data; (ii) python/qsymm
  \cite{Qsymm2018} implementation of the symmetry constraints to obtain
  graphene's effective models; (iii) python/kwant \cite{kwant} implementation
  of graphene's tight-binding model; and (iv) numerical (python) implementation
  of the effective models using finite differences, and the codes used to
  generate the figures.}}\BibitemShut {Stop}%
\bibitem [{\citenamefont {Perdew}\ \emph {et~al.}(1996)\citenamefont {Perdew},
  \citenamefont {Burke},\ and\ \citenamefont {Ernzerhof}}]{GGAPBE35}%
  \BibitemOpen
  \bibfield  {author} {\bibinfo {author} {\bibfnamefont {J.~P.}\ \bibnamefont
  {Perdew}}, \bibinfo {author} {\bibfnamefont {K.}~\bibnamefont {Burke}},\ and\
  \bibinfo {author} {\bibfnamefont {M.}~\bibnamefont {Ernzerhof}},\ }\bibfield
  {title} {\bibinfo {title} {{Generalized Gradient Approximation Made
  Simple}},\ }\href {https://doi.org/10.1103/PhysRevLett.77.3865} {\bibfield
  {journal} {\bibinfo  {journal} {Phys. Rev. Lett.}\ }\textbf {\bibinfo
  {volume} {77}},\ \bibinfo {pages} {3865} (\bibinfo {year}
  {1996})}\BibitemShut {NoStop}%
\bibitem [{\citenamefont {Bl\"ochl}(1994)}]{PAW36}%
  \BibitemOpen
  \bibfield  {author} {\bibinfo {author} {\bibfnamefont {P.~E.}\ \bibnamefont
  {Bl\"ochl}},\ }\bibfield  {title} {\bibinfo {title} {{Projector
  augmented-wave method}},\ }\href {https://doi.org/10.1103/PhysRevB.50.17953}
  {\bibfield  {journal} {\bibinfo  {journal} {Phys. Rev. B}\ }\textbf {\bibinfo
  {volume} {50}},\ \bibinfo {pages} {17953} (\bibinfo {year}
  {1994})}\BibitemShut {NoStop}%
\bibitem [{\citenamefont {Kresse}\ and\ \citenamefont
  {Furthm\"uller}(1996)}]{VASP33}%
  \BibitemOpen
  \bibfield  {author} {\bibinfo {author} {\bibfnamefont {G.}~\bibnamefont
  {Kresse}}\ and\ \bibinfo {author} {\bibfnamefont {J.}~\bibnamefont
  {Furthm\"uller}},\ }\bibfield  {title} {\bibinfo {title} {{Efficiency of
  ab-initio total energy calculations for metals and semiconductors using a
  plane-wave basis set}},\ }\href
  {https://doi.org/10.1016/0927-0256(96)00008-0} {\bibfield  {journal}
  {\bibinfo  {journal} {Comput. Mater. Sci.}\ }\textbf {\bibinfo {volume}
  {6}},\ \bibinfo {pages} {15} (\bibinfo {year} {1996})}\BibitemShut {NoStop}%
\bibitem [{\citenamefont {Kresse}\ and\ \citenamefont
  {Furthmüller}(1996)}]{VASP34}%
  \BibitemOpen
  \bibfield  {author} {\bibinfo {author} {\bibfnamefont {G.}~\bibnamefont
  {Kresse}}\ and\ \bibinfo {author} {\bibfnamefont {J.}~\bibnamefont
  {Furthmüller}},\ }\bibfield  {title} {\bibinfo {title} {{Efficient iterative
  schemes for ab initio total-energy calculations using a plane-wave basis
  set}},\ }\href {https://doi.org/10.1103/PhysRevB.54.11169} {\bibfield
  {journal} {\bibinfo  {journal} {Phys. Rev. B}\ }\textbf {\bibinfo {volume}
  {54}},\ \bibinfo {pages} {11169} (\bibinfo {year} {1996})}\BibitemShut
  {NoStop}%
\bibitem [{\citenamefont {Winkler}(2003)}]{winkler2003spin}%
  \BibitemOpen
  \bibfield  {author} {\bibinfo {author} {\bibfnamefont {R.}~\bibnamefont
  {Winkler}},\ }\href {https://doi.org/10.1007/b13586} {\emph {\bibinfo {title}
  {Spin-orbit coupling effects in two-dimensional electron and hole
  systems}}},\ \bibinfo {series} {Springer Tracts in Modern Physics}, Vol.\
  \bibinfo {volume} {191}\ (\bibinfo  {publisher} {Springer-Verlag, Berlin,
  Heidelberg},\ \bibinfo {year} {2003})\BibitemShut {NoStop}%
\bibitem [{\citenamefont {Alonso}\ \emph {et~al.}(1997)\citenamefont {Alonso},
  \citenamefont {{De Vincenzo}},\ and\ \citenamefont
  {Mondino}}]{Alonso1997DiracBC2}%
  \BibitemOpen
  \bibfield  {author} {\bibinfo {author} {\bibfnamefont {V.}~\bibnamefont
  {Alonso}}, \bibinfo {author} {\bibfnamefont {S.}~\bibnamefont {{De
  Vincenzo}}},\ and\ \bibinfo {author} {\bibfnamefont {L.}~\bibnamefont
  {Mondino}},\ }\bibfield  {title} {\bibinfo {title} {On the boundary
  conditions for the dirac equation},\ }\href
  {https://doi.org/10.1088/0143-0807/18/5/001} {\bibfield  {journal} {\bibinfo
  {journal} {Eur. J. Phys.}\ }\textbf {\bibinfo {volume} {18}},\ \bibinfo
  {pages} {315} (\bibinfo {year} {1997})}\BibitemShut {NoStop}%
\bibitem [{\citenamefont {Alonso}\ and\ \citenamefont {{De
  Vincenzo}}(1997)}]{Alonso1997DiracBC}%
  \BibitemOpen
  \bibfield  {author} {\bibinfo {author} {\bibfnamefont {V.}~\bibnamefont
  {Alonso}}\ and\ \bibinfo {author} {\bibfnamefont {S.}~\bibnamefont {{De
  Vincenzo}}},\ }\bibfield  {title} {\bibinfo {title} {General boundary
  conditions for a dirac particle in a box and their non-relativistic limits},\
  }\href {https://doi.org/10.1088/0305-4470/30/24/018} {\bibfield  {journal}
  {\bibinfo  {journal} {J. Phys. A: : Math. Gen.}\ }\textbf {\bibinfo {volume}
  {30}},\ \bibinfo {pages} {8573} (\bibinfo {year} {1997})}\BibitemShut
  {NoStop}%
\bibitem [{\citenamefont {Ferreira}\ and\ \citenamefont
  {Loss}(2013)}]{Ferreira2013Magnetically}%
  \BibitemOpen
  \bibfield  {author} {\bibinfo {author} {\bibfnamefont {G.~J.}\ \bibnamefont
  {Ferreira}}\ and\ \bibinfo {author} {\bibfnamefont {D.}~\bibnamefont
  {Loss}},\ }\bibfield  {title} {\bibinfo {title} {{Magnetically Defined Qubits
  on 3D Topological Insulators}},\ }\href
  {https://doi.org/10.1103/PhysRevLett.111.106802} {\bibfield  {journal}
  {\bibinfo  {journal} {Phys. Rev. Lett.}\ }\textbf {\bibinfo {volume} {111}},\
  \bibinfo {pages} {106802} (\bibinfo {year} {2013})}\BibitemShut {NoStop}%
\bibitem [{mas()}]{massRange}%
  \BibitemOpen
  \bibinfo {note} {{In the continuum limit $\delta_x \rightarrow 0$.
  Additionally, requiring that the linear spectrum to dominate over the full
  energy range $\delta_\varepsilon \rightarrow \infty$, the constraints in
  Eq.~\eqref{eq:mass} yields $m \equiv 0$.}}\BibitemShut {Stop}%
\bibitem [{\citenamefont {Son}\ \emph {et~al.}(2006)\citenamefont {Son},
  \citenamefont {Cohen},\ and\ \citenamefont {Louie}}]{Son2006DFTgap}%
  \BibitemOpen
  \bibfield  {author} {\bibinfo {author} {\bibfnamefont {Y.-W.}\ \bibnamefont
  {Son}}, \bibinfo {author} {\bibfnamefont {M.~L.}\ \bibnamefont {Cohen}},\
  and\ \bibinfo {author} {\bibfnamefont {S.~G.}\ \bibnamefont {Louie}},\
  }\bibfield  {title} {\bibinfo {title} {Energy gaps in graphene nanoribbons},\
  }\href {https://doi.org/10.1103/PhysRevLett.97.216803} {\bibfield  {journal}
  {\bibinfo  {journal} {Phys. Rev. Lett.}\ }\textbf {\bibinfo {volume} {97}},\
  \bibinfo {pages} {216803} (\bibinfo {year} {2006})}\BibitemShut {NoStop}%
\bibitem [{\citenamefont {DiVincenzo}\ and\ \citenamefont
  {Mele}(1984)}]{DiVincenzo1984GrapheneKP}%
  \BibitemOpen
  \bibfield  {author} {\bibinfo {author} {\bibfnamefont {D.~P.}\ \bibnamefont
  {DiVincenzo}}\ and\ \bibinfo {author} {\bibfnamefont {E.~J.}\ \bibnamefont
  {Mele}},\ }\bibfield  {title} {\bibinfo {title} {Self-consistent
  effective-mass theory for intralayer screening in graphite intercalation
  compounds},\ }\href {https://doi.org/10.1103/PhysRevB.29.1685} {\bibfield
  {journal} {\bibinfo  {journal} {Phys. Rev. B}\ }\textbf {\bibinfo {volume}
  {29}},\ \bibinfo {pages} {1685} (\bibinfo {year} {1984})}\BibitemShut
  {NoStop}%
\bibitem [{\citenamefont {Bastard}(1988)}]{BastardBook}%
  \BibitemOpen
  \bibfield  {author} {\bibinfo {author} {\bibfnamefont {G.}~\bibnamefont
  {Bastard}},\ }\href@noop {} {\emph {\bibinfo {title} {{Wave mechanics applied
  to semiconductor heterostructures}}}},\ Monographies de physique\ (\bibinfo
  {publisher} {Les {\'E}ditions de Physique},\ \bibinfo {year}
  {1988})\BibitemShut {NoStop}%
\bibitem [{\citenamefont {Ando}(2005)}]{Ando2005ReviewNanotubes}%
  \BibitemOpen
  \bibfield  {author} {\bibinfo {author} {\bibfnamefont {T.}~\bibnamefont
  {Ando}},\ }\bibfield  {title} {\bibinfo {title} {Theory of electronic states
  and transport in carbon nanotubes},\ }\href
  {https://doi.org/10.1143/JPSJ.74.777} {\bibfield  {journal} {\bibinfo
  {journal} {J. Phys. Soc. Jpn.}\ }\textbf {\bibinfo {volume} {74}},\ \bibinfo
  {pages} {777} (\bibinfo {year} {2005})}\BibitemShut {NoStop}%
\bibitem [{\citenamefont {Ara\'ujo}\ \emph {et~al.}(2016)\citenamefont
  {Ara\'ujo}, \citenamefont {Wrasse}, \citenamefont {Ferreira},\ and\
  \citenamefont {Schmidt}}]{Araujo2016Nonsymm}%
  \BibitemOpen
  \bibfield  {author} {\bibinfo {author} {\bibfnamefont {A.~L.}\ \bibnamefont
  {Ara\'ujo}}, \bibinfo {author} {\bibfnamefont {E.~O.}\ \bibnamefont
  {Wrasse}}, \bibinfo {author} {\bibfnamefont {G.~J.}\ \bibnamefont
  {Ferreira}},\ and\ \bibinfo {author} {\bibfnamefont {T.~M.}\ \bibnamefont
  {Schmidt}},\ }\bibfield  {title} {\bibinfo {title} {Topological nonsymmorphic
  ribbons out of symmorphic bulk},\ }\href
  {https://doi.org/10.1103/PhysRevB.93.161101} {\bibfield  {journal} {\bibinfo
  {journal} {Phys. Rev. B}\ }\textbf {\bibinfo {volume} {93}},\ \bibinfo
  {pages} {161101(R)} (\bibinfo {year} {2016})}\BibitemShut {NoStop}%
\bibitem [{\citenamefont {Wakabayashi}\ \emph {et~al.}(2010)\citenamefont
  {Wakabayashi}, \citenamefont {i.~Sasaki}, \citenamefont {Nakanishi},\ and\
  \citenamefont {Enoki}}]{Wakabayashi2012Density}%
  \BibitemOpen
  \bibfield  {author} {\bibinfo {author} {\bibfnamefont {K.}~\bibnamefont
  {Wakabayashi}}, \bibinfo {author} {\bibfnamefont {K.}~\bibnamefont
  {i.~Sasaki}}, \bibinfo {author} {\bibfnamefont {T.}~\bibnamefont
  {Nakanishi}},\ and\ \bibinfo {author} {\bibfnamefont {T.}~\bibnamefont
  {Enoki}},\ }\bibfield  {title} {\bibinfo {title} {Electronic states of
  graphene nanoribbons and analytical solutions},\ }\href
  {https://doi.org/10.1088/1468-6996/11/5/054504} {\bibfield  {journal}
  {\bibinfo  {journal} {Sci. Technol. Adv. Mater}\ }\textbf {\bibinfo {volume}
  {11}},\ \bibinfo {pages} {054504} (\bibinfo {year} {2010})}\BibitemShut
  {NoStop}%
\bibitem [{\citenamefont {Liu}\ \emph {et~al.}(2015)\citenamefont {Liu},
  \citenamefont {Qian},\ and\ \citenamefont {Fu}}]{PbSe2}%
  \BibitemOpen
  \bibfield  {author} {\bibinfo {author} {\bibfnamefont {J.}~\bibnamefont
  {Liu}}, \bibinfo {author} {\bibfnamefont {X.}~\bibnamefont {Qian}},\ and\
  \bibinfo {author} {\bibfnamefont {L.}~\bibnamefont {Fu}},\ }\bibfield
  {title} {\bibinfo {title} {{Crystal Field Effect Induced Topological
  Crystalline Insulators In Monolayer IV–VI Semiconductors}},\ }\href
  {https://doi.org/10.1021/acs.nanolett.5b00308} {\bibfield  {journal}
  {\bibinfo  {journal} {Nano Letters}\ }\textbf {\bibinfo {volume} {15}},\
  \bibinfo {pages} {2657} (\bibinfo {year} {2015})}\BibitemShut {NoStop}%
\bibitem [{\citenamefont {Niu}\ \emph {et~al.}(2015)\citenamefont {Niu},
  \citenamefont {Buhl}, \citenamefont {Bihlmayer}, \citenamefont {Wortmann},
  \citenamefont {Bl\"ugel},\ and\ \citenamefont {Mokrousov}}]{PbSe3}%
  \BibitemOpen
  \bibfield  {author} {\bibinfo {author} {\bibfnamefont {C.}~\bibnamefont
  {Niu}}, \bibinfo {author} {\bibfnamefont {P.~M.}\ \bibnamefont {Buhl}},
  \bibinfo {author} {\bibfnamefont {G.}~\bibnamefont {Bihlmayer}}, \bibinfo
  {author} {\bibfnamefont {D.}~\bibnamefont {Wortmann}}, \bibinfo {author}
  {\bibfnamefont {S.}~\bibnamefont {Bl\"ugel}},\ and\ \bibinfo {author}
  {\bibfnamefont {Y.}~\bibnamefont {Mokrousov}},\ }\bibfield  {title} {\bibinfo
  {title} {{Topological crystalline insulator and quantum anomalous Hall states
  in IV-VI-based monolayers and their quantum wells}},\ }\href
  {https://doi.org/10.1103/PhysRevB.91.201401} {\bibfield  {journal} {\bibinfo
  {journal} {Phys. Rev. B}\ }\textbf {\bibinfo {volume} {91}},\ \bibinfo
  {pages} {201401(R)} (\bibinfo {year} {2015})}\BibitemShut {NoStop}%
\bibitem [{not()}]{noteBCprev}%
  \BibitemOpen
  \bibinfo {note} {{In Ref.~\cite{Araujo2016Nonsymm}, we have considered a
  Brey-Fertig type of boundary condition applied at the different atomic
  terminations of each sublattice (Pb/Se). This gives the correct results, as
  it breaks chiral symmetry properly for ribbons A and B. Within our approach,
  the same result is much simpler to obtain with a finite $\rho$.}}\BibitemShut
  {Stop}%
\end{thebibliography}%

\end{document}